\documentclass[showpacs,preprintnumbers,amsmath,twocolumn,amssymb,revsymb, superscriptaddress,aps]{revtex4-1}
\usepackage{mathtools}
\usepackage{graphicx}
\usepackage[T1]{fontenc}
\usepackage{amsmath, amsthm, amssymb}
\usepackage{amssymb}

\usepackage{xcolor}
\usepackage{appendix}

\newcommand{\tu}{\tilde{u}}
\newcommand{\tv}{\tilde{v}}
\newcommand{\hu}{\hat{u}}
\newcommand{\hv}{\hat{v}}

\def\ketbra#1{|#1\rangle\langle#1|}
\def\tr{\mathrm{tr}}

\DeclarePairedDelimiter\bra{\langle}{\rvert}
\DeclarePairedDelimiter\ket{\lvert}{\rangle}
\DeclarePairedDelimiterX\braket[2]{\langle}{\rangle}{#1 \delimsize\vert #2}

\begin{document}
\title{ Quantum Hamiltonian Computing protocols for molecular electronics Boolean logic gates}

\author{Omid Faizy Namarvar } 
\affiliation{ Laboratoire de Physique Th\'eorique, IRSAMC, Universit\'e de Toulouse, CNRS, UPS, France}
\affiliation{ CEMES-CNRS, 29 rue J. Marvig, 31055 Toulouse Cedex,France}

\author{Olivier Giraud } 
\affiliation{ LPTMS, CNRS, Univ. Paris-Sud, Universit\'e Paris-Saclay - 91405 Orsay, France}

\author{Bertrand Georgeot} 
\affiliation{ Laboratoire de Physique Th\'eorique, IRSAMC, Universit\'e de Toulouse, CNRS, UPS, France}

\author{Christian Joachim }  
\affiliation{ CEMES-CNRS, 29 rue J. Marvig, 31055 Toulouse Cedex,France}
\affiliation{ WPI-MANA, National Institute for Material Sciences, 1-1 Namiki, Tsukuba, Ibaraki, Japan}

\date{\today}

\begin{abstract}
Quantum Hamiltonian Computing is a recent approach that uses quantum systems, in particular a single molecule, to perform computational tasks. Within this approach, we present explicit methods to construct logic gates using two different designs, where the logical outputs are encoded either at fixed energy and spatial positioning of the quantum states, or at different energies.  We use these results to construct quantum Boolean adders involving a minimal number of quantum states with the two designs. We also establish a matrix algebra giving an analogy between classical Boolean logic gates and quantum ones, and assess the possibilities of both designs for more complex gates.  

\end{abstract}


\maketitle


\section{Introduction}

The use of quantum mechanics to treat information has been the subject of an enormous amount of works in the recent past. In particular, quantum computing promises to create computers of a new type, which could change the complexity class of certain problems \cite{refqc}. In most cases, the design is based on elementary two-level systems (qubits) on which one could act through quantum gates to perform sequences of operations (quantum circuit model). This enables to run quantum algorithms outperforming classical ones, like the Shor \cite{Shor} and Grover \cite{Grover} algorithms. Many systems have been proposed to be building blocks of such a device, from e.g. trapped ions to superconducting mesoscopic devices.

A parallel stream of works focused on molecular systems, trying to build elementary logical operations using the versatility and controllability of a single molecule. After the seminal work of A. Aviram and M. Ratner \cite{aviram}, a few hybrid molecular electronic circuits have been proposed \cite{Gimzewski}. The semi-classical mono-molecular approach was then introduced by \cite{carter} where the entire arithmetic and logic unit (ALU) of a calculator was proposed to be embedded in a single very large molecule (Y. Wada proposed the same with electronic circuits supposed to be constructed atom by atom on a surface \cite{wada}). A more chemically realistic single-molecule 2-digit full adder was proposed by J. Ellenbogen \cite{Ellenbogen}. There the design rules were based on the G. Kirchhoff meshes and nodes circuit rules. It was demonstrated later to be unrealistic, taking into account the quantum electron transfer processes occurring through a molecule \cite{magoga}. As a consequence, pushing a large molecule to have the shape of an electrical circuit and to function like a classical electrical Boolean logic circuit may not be the way to go for miniaturizing the ALU down to the atomic scale, for example inside a single molecule.

In this {\it a la} Shannon architecture for molecular electronic circuits, in order to obtain complex Boolean functions, multiple layers of electronic logic gates such as OR, NOT, NAND, NOR in a planar configuration are required. By contrast, quantum computation based on the implementation of quantum logic gates relies on the manipulation of quantum states inside a quantum electronic system, such as a single conjugated molecule or a few atoms stabilized in a cold atomic trap. Within a quantum circuit, it is a unitary transformation, implemented via a time-dependent quantum evolution, which performs the computation. 

In order to perform quantum computing inside a molecule, the first idea was to introduce well-separated qubits along the molecular structure. This approach was first experimented by I. Chuang in 1998 using NMR techniques \cite{chaung}. But although a proof of concept was obtained, these quantum gates cannot be miniaturized down to the atomic scale. By revisiting how quantum control works on such systems, a new quantum control protocol, called the Quantum Hamiltonian Computing approach (QHC) was proposed in 2005 \cite{Duchemin }. It does not involve dividing the molecular structure into individual qubits, but it still belongs to the same family of control theory than the one generally applied to qubit systems. This similarity has enabled a cross-fertilization between the qubit and the new QHC approaches since they both exploit quantum superposition respecting normalization (Born principle) \cite{joachim1}. A similar idea was also recently proposed in \cite{aspuru}. 

The basic principle of QHC is the following. An arbitrary logical calculation takes as input a string $\{\alpha_1, \cdots \alpha_k \}$ of 0 or 1, and its output is a string $\{ \mu_1, \cdots \mu_l\}$ of 0 and 1. Let us consider a quantum system prepared in an initial state $\ket{\psi(0)}$ and undergoing a time evolution governed by some Hamiltonian $H$, so that at time $t$ the system is in a state $\ket{\psi(t)}=e^{-iHt/\hbar} \ket{\psi(0)}$. It is possible to map any logical operation to be performed on the input $\{\alpha_1, \cdots \alpha_k \}$ onto a quantum trajectory of the time-dependent state vector $\ket{\psi(t)}$ \cite{renaud1}. This mapping depends on the practical way chosen to encode the logical inputs on the quantum system and on the procedure to measure the logical outputs. For example, the inputs can be encoded either into the initial state $\ket{\psi(0)}$ (as in the qubit approach) or into matrix elements of the Hamiltonian $H$ governing the evolution (as is the case in the QHC approach). 

In the situation where the molecule is divided in qubits spatially distributed over the molecular structure, as in \cite{chaung}, then $\ket{\psi(0)}=\ket{ \alpha_1, \cdots \alpha_k}$ carries the logical inputs, while $H$ is in general independent of the logical input configuration and runs the quantum evolution. A target state vector $\ket{\varphi}= \ket{\mu_1,\cdots\mu_l}$ is then associated with any output string, and specific characteristics of the $\braket{\varphi} {\psi(t)}$ population amplitude, such as the $|\braket{\varphi} {\psi(t)}|^2$ maximum over time, can be used to define an appropriate output measurement strategy. These measurements have to be performed before decoherence of the $\ket{\psi(t)}$ wavepacket sets in, and also before it relaxes to the ground state of the quantum system. The Hamiltonian should perform a unitary transformation $\mathcal{B} \ket{ \alpha_1, \cdots \alpha_k }= \ket{ \mu_1, \cdots \mu_l }$, with $\mathcal{B}$ constructed in such a way that the results of the Boolean calculations are measurable exactly at specific times $t_n$ (in such a qubit design, the $\ket{\psi(t)}\bra{\psi(t)}$ quantum trajectory is normally fully periodic \cite{chaung,hliwa}).

By contrast, in the QHC approach that we will consider here, the input string $ \{ \alpha_1, \cdots \alpha_k \}$ is now encoded into the Hamiltonian generating the quantum time evolution, so that $H=H( \alpha_1, \cdots \alpha_k )$. The $ \{\mu_1, \cdots \mu_l\}$  output string is then measured using well-selected ''pointer'' states. There are two different ways of achieving the measurement. One possibility is to fix an initial state $\ket{\psi(0)}=\ket{\varphi_a}$ and a pointer state $\ket{\varphi_b}$ per output bit, and to attribute a logical output '1' when the quantum trajectory $\ket{\psi(t)}=e^{-iHt/\hbar} \ket{\varphi_a}$ reaches the target state $\ket{\varphi_b}$ at $t = t_n$ [11]. However this is not very practical for an atomic scale implementation. A second possibility, which will be the one considered in the present paper, is to choose a pair $\ket{\varphi_a}$, $\ket{\varphi_b}$ of pointer states per output bit, and to encode that bit into the value of the $\Omega_{ab}$ secular oscillation frequency between $\ket{\varphi_a}$ and $\ket{\varphi_b}$, which depends on $H$ and thus on the input. For instance, a very fast oscillation will encode for an output '1' and a very slow oscillation for a '0'. A very important advantage of this output encoding is that the tunneling current intensity passing through the measurement point of the molecule is proportional to $\Omega_{ab}^2$ (this was verified experimentally with the starphene molecule \cite{skidin,Soe}).  

In this paper, QHC Hamiltonians are constructed to perform Boolean half adders and Boolean full adders. Half adders were already constructed heuristically  \cite{dridi1, dridi2}, but no full adder has been obtained before. We present a systematic approach in order to reach more complex logical operations with a minimum of quantum states, obtaining explicit examples of full adders. Depending on the complexity, two different output measurement strategies will be deployed. For small Boolean logic gates, the logical output will be measured on different states of the QHC quantum system at the same energy as in \cite{dridi1}. For more complex Boolean logic gates, it tuns out to be preferable that each output digit has its own different reading energy, and the reading may occur on the same QHC quantum state or not. In Sec \ref{booleanalgebra}, we recall and complete the interpretation of our preceding designs concerning this first output strategy starting from the simplest 3 states calculating block Hamiltonian. We reinterpret them in terms of a procedure merging simple gates into more complex ones at the quantum level. In Sec \ref{generalization}, we present a systematic approach to construct more complex QHC logic gates Hamiltonians along this output strategy. We use the secular frequency $\Omega_{ab}$ of the the Heisenberg-Rabi time-dependent oscillations through the calculating block to characterize the functioning of those gates. We present examples of half adder Hamiltonians and the first example of a full adder with this approach. We then use the systematic approach to study the feasibility of more complex gates. To go further in complexity with a minimum of quantum states in the calculating block, we present in Sec \ref{scanning} the QHC Hamiltonian optimization for multiple energy reading blocks. In this case, we use the transmission coefficient \cite{sautet} through the QHC calculating block (proportional to $\Omega_{ab}^2$ per reading block \cite{faizy}) to interpret the multi energy reading. We present a protocol to build quantum gates with this design, and use it to construct a half adder and a full adder. We also explore the possibilities for larger gates using this design.

\section{Boolean algebra approach}
\label{booleanalgebra}

\subsection{The basic principle}
In QHC, each logical output of a gate is determined by measuring either the oscillation frequency $\Omega_{ab}$ or the tunneling current intensity between two nearby locations on the QHC molecule which is proportional to $\Omega_{ab}^2$. The two states $\ket{\varphi_a}$ and $\ket{\varphi_b}$ define the ''reading block'' for this output and are generally weakly coupled via a coupling constant $\varepsilon\ll 1$ to the molecule, which is called hereafter the ''calculating block''.

Since each reading block quantum state subspace is spanned by $\ket{\varphi_a}$ and $\ket{\varphi_b}$ of energy $E$, the control of the calculating block on the Heisenberg-Rabi oscillations can be described for small $\varepsilon\ll 1$ by the $2\times 2$ effective Hamiltonian \cite{PhysChem} 

\begin{equation}
H_{\textrm{eff}}=\frac{\hbar}{2}
\left(
\begin{array}{cc}
 0 &\Omega_{ab}\\
\Omega_{ab}&0
\end{array}
\right),
\end{equation}

with $\Omega_{ab}$ the secular oscillating frequency between $\ket{\varphi_a}$ and $\ket{\varphi_b}$ through the calculating block, given by

\begin{equation}
\label{offdiag}
\frac{\hbar\Omega_{ab}}{2}=\epsilon^2\lim_{\eta\to 0}\tr\left(P\frac{1}{E-H_0+i\eta}\right),
\end{equation}
where $P=\ketbra{\varphi_a}+\ketbra{\varphi_b}$ is the projector on the reading block subspace and $H_0$ is the Hamiltonian of the calculating block. According to \eqref{offdiag}, if for some input values, an eigenvalue of $H_0$ takes the value $E$ and the corresponding eigenvector has a nonzero projection on $\{\ket{\varphi_a},\ket{\varphi_b}\}$ then $\Omega_{ab}$ becomes large. If one of these two conditions is not met, $\Omega_{ab}$ will be very small. For QHC, a large $\Omega_{ab}$ codes for logical output '1' and a very small one for logical output '0'. The goal of this section is to recall, first on simple Boolean gates, how to construct $H_0$ depending on the input string $\{\alpha_1, \cdots \alpha_k\}$ in such a way that $\Omega_{ab}$ is large for output 1 and very small for output 0.

\subsection{The Boolean QHC gate Hamiltonian}

The essence of the QHC method is best illustrated on the simple case of the two-input/one-output symmetric Boolean logic gates AND, OR, XOR, NAND, NOR, NXOR. The logical output of the gates considered here is given by its usual Boolean expression, recalled in the second column of Table \ref{gates}. The symbols $\lor$, $\land$ and $\lnot$ stand for the logical OR and AND and NOT respectively leading to $\lnot(\alpha\lor\beta)=\lnot\alpha\land\lnot\beta$ and $\alpha\oplus\beta=(\alpha\land\lnot\beta)\lor(\beta\land\lnot\alpha)$. Replacing the logical 0 and 1 by the numbers 0 and 1, these logical operations can me mapped to simple algebraic operations in the discrete ring $\mathbb{Z}/2\mathbb{Z}\equiv \{0,1\}$, using the correspondence $\alpha\lor\beta \equiv \alpha+\beta$ for the OR and $\alpha\land\beta=\alpha\beta$ for the AND, while the NOT corresponds for QHC to $\lnot\alpha\equiv 1-\alpha$. If the two inputs are $\alpha,\beta\in\{0,1\}$, this correspondence allows to express the gate output $\mu(\alpha,\beta)\in\{0,1\}$ as a function of the logical input numbers, as displayed in the third column of Table \ref{gates}. This correspondence is adapted to the QHC approach since there is no quantum gain possible from inside a single molecule or an atomic scale circuit (according to the Born principle), meaning that the NOT must be obtained by playing with time dependent destructive (constructive) interferences.

\begin{figure}[!b]
    \centering
    \includegraphics[width=.6\linewidth]{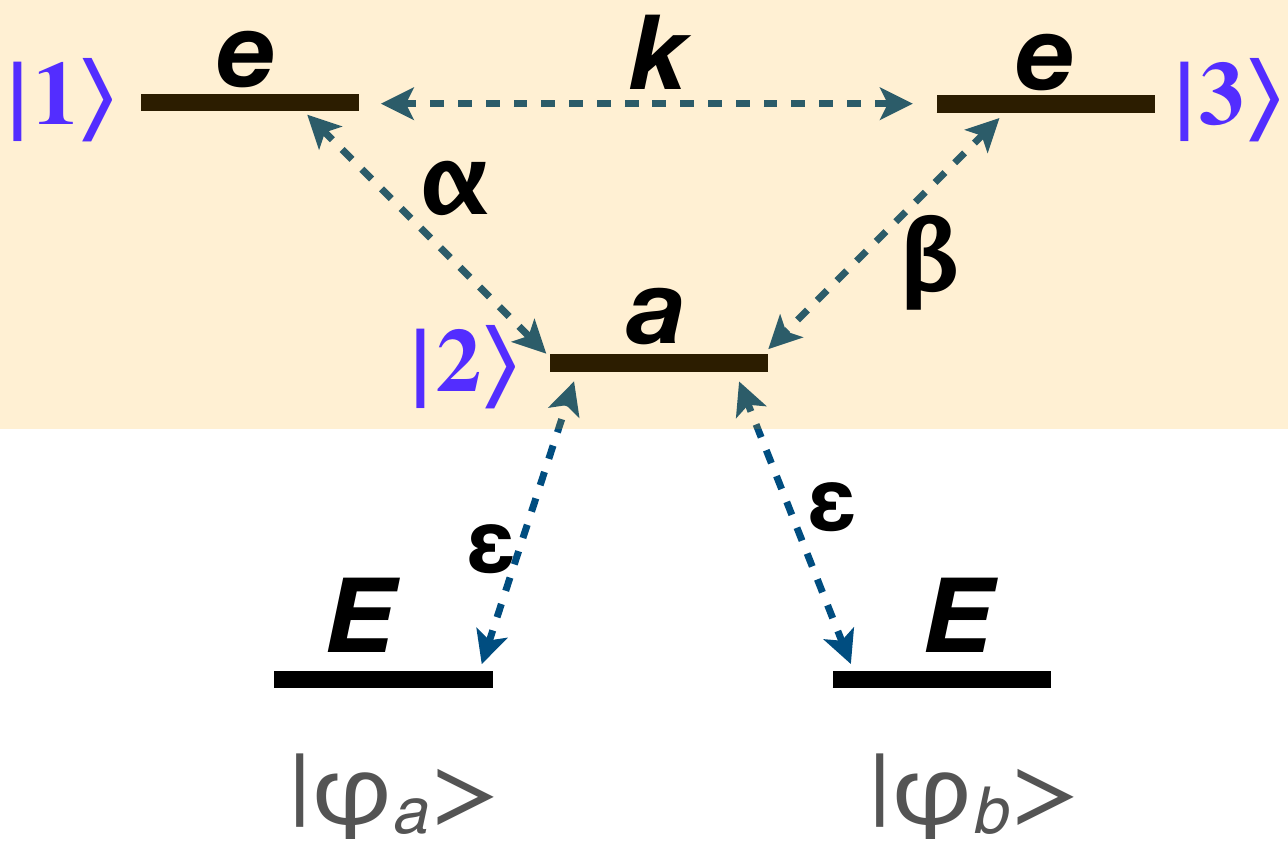} 
   \caption{Quantum graph representation of Hamiltonian \eqref{eq:hamiltonianH1} performing all six symmetric two-input/one-output Boolean logic gates. The quantum graph of the calculating part $H_0$ of $H$ is shaded.}
  \label{fig:3_states}
  \end{figure}

\begin{table*}[!t]
\begin{center}
   \caption{  \label{gates}
The six elementary Boolean logic gates constructed from the $3\times 3$ calculating block given by Eq.(\refeq{eq:hamiltonianH0})}
\begin{tabular}{|c|c|c|c|c|c|}
Gate & Boolean algebra  &$\mu(\alpha,\beta)$ (boolean ring) & Parameters & $P_{\alpha\beta}(0)$ & Non-boolean $P_{\alpha\beta}(0)$\\
\hline
AND &$\alpha\land\beta$  &$\alpha\beta$ & $e=0$, $a=\frac{2}{k}$ &  $-2k(1-\alpha\beta)$ & $2-\alpha^2-\beta^2$\\
OR &$\alpha\lor\beta$  & $\alpha+\beta-\alpha\beta$ & $e=2k$, $a=\frac{2}{3k}$ &  $2k(1-\alpha-\beta+\alpha\beta)$ &$1-\alpha-\beta+\alpha\beta$ \\
XOR &$\alpha\oplus\beta$  & $\alpha+\beta-2\alpha\beta$& no solution & no solution & $1-\alpha^2-\beta^2$\\
NAND & $\lnot(\alpha\land\beta)$ &$1-\alpha\beta$&  $e=a=0$   &$2k\alpha\beta$  &no solution  \\
NOR &$\lnot(\alpha\lor\beta)$  &$1-\alpha-\beta+\alpha\beta$& $e=\frac{1}{2k}$, $a=0$   &$e(\alpha+\beta-\alpha\beta)$& $\alpha^2+\beta^2$\\
NXOR &$\lnot(\alpha\oplus\beta)$   &$1-\alpha-\beta+2\alpha\beta$&  $e=k$  & $-k(\alpha+\beta-2\alpha\beta)$ & no solution
\end{tabular}
\end{center}
 \label{Table}
\end{table*}

As demonstrated recently in \cite{dridi2}, the calculating block described by $H_0$ should at least be $3\times 3$ in order to design all 6 logic gates. The calculating block consists of two states of energy $e$ coupled via $\alpha$ and $\beta$ to a state of energy $a$, see Fig.~\ref{fig:3_states}. This latter state is going to be shifted in energy as a function of the logical input values, together with the complete $H_0$ eigenspectrum. The Hamiltonian $H_0$ describing this calculating block has the minimal $3\times 3$ matrix form:
\begin{equation} \label{eq:hamiltonianH0}
H_0(\alpha,\beta)= \left[ 
    \begin{matrix}
      e & \alpha & k  \\
          \alpha &a &  \beta \\
            k & \beta & e
      \end{matrix}
    \right].      
\end{equation}
Here, $a$, $e$, and $k$ are the structural parameters whose tuning allows to build the gates.

In order to read the logical output, the $a$ state is weakly electronically coupled via the small term $\varepsilon$ to the reading block, as presented in Fig.~\ref{fig:3_states}. Since all 6 logic gates considered here are symmetric with respect to permutation of $\alpha$ and $\beta$, the choice of coupling the reading block to level $a$ directly ensures that this symmetry is fulfilled. If the two pointer states $\ket{\varphi_a}$ and $\ket{\varphi_b}$ are at energy $E$, the complete QHC logic gate Hamiltonian reads

\begin{equation}
\label{eq:hamiltonianH1}
H(\alpha,\beta,E)= \left[ 
    \begin{matrix}
      e & \alpha & k &0&0 \\
          \alpha &a &  \beta&\varepsilon&\varepsilon\\
            k & \beta & e &0&0\\
            0&\varepsilon&0&E&0\\
            0&\varepsilon&0&0&E
      \end{matrix}
    \right].      
\end{equation}

For a fixed set of structural parameters $a$, $e$ and $k$, the complete system is prepared at $t=0$ in state $\ket{\varphi_a}$. From $t=0$, it will spontaneously oscillate between state $\ket{\varphi_a}$ and state $\ket{\varphi_b}$ through the calculating block states at a frequency $\Omega_{ab}$ given at lowest order in $\varepsilon$ by \eqref{offdiag}. Depending on the input $(\alpha,\beta)$, one eigenstate of the calculating block may resonate with $\ket{\varphi_a}$ and $\ket{\varphi_b}$. In that case, it will result in a fast and alternating Heisenberg-Rabi time-dependent evolution of the $\ket{\varphi_a}$ and $\ket{\varphi_b}$ occupation. The coupling constant $\varepsilon$ must be small relative to the difference taken two by two between the three eigenvalues of the calculating block in order not to perturb this time-dependent evolution  \cite{dridi2}. Note that it is also important to have a good control on the $\alpha$ and $\beta$ logical input changes in time. When their variations are faster than the secular frequency of the Heisenberg-Rabi oscillations, the $H_0$ eigenvalues responsible for the secular oscillations will change before the ramping up in time of those oscillations. This is a net difference with the qubit approach, which is not separating the calculating from the reading time sequences.

\subsection{Finding the structural parameters}
There are two non-exclusive choices to determine the reading energy $E$ in \eqref{eq:hamiltonianH1}. Described in this section, the first choice is to fix $E$ whatever the logic gate truth table and then determine the structural parameters as a function of the targeted truth table. The other choice is to determine a set of structural parameters common to all the gates (which for example are imposed by the atomic or molecular scale implementation of the QHC gate) and then read the logical outputs at different energies $E$, one per logic gate truth table \cite{kolmer}. This second choice will be described in Sec \ref{scanning}.

The calculating block is determined using the characteristic polynomial of $H_0$:
\begin{equation}
\label{charpolH0}
P_{\alpha\beta}(E)=(e-E)^{2}(a-E)-(e-E)(\alpha^{2}+\beta^{2})-k^{2}(a-E)+2k\alpha\beta.
\end{equation}

When the reading energy $E$ is fixed, the logical output 1 will be reached for a certain logical input $\alpha,\beta$ only if some eigenvalue of $H_0(\alpha,\beta)$ is equal to $E$. According to \eqref{charpolH0}, this means that the structural parameters must be selected in such a way that $P_{\alpha\beta}(E)=0$ whenever the output $\mu(\alpha,\beta)$ is 1, and $P_{\alpha\beta}(E)$ is nonzero whenever $\mu(\alpha,\beta)=0$. The simplest way of achieving that is to take $P_{\alpha\beta}(E) \propto 1-\mu(\alpha,\beta)$. Imposing that relation allows to find the parameters $a$, $e$ and $k$ for the targeted Boolean truth table. 

The case $E=0$ for all the 2-inputs/1-output gates was studied recently \cite{renaud1}. From  \eqref{charpolH0} we have $P_{\alpha\beta}(0)=a \left(e^2-k^2\right)-e (\alpha +\beta )+2 \alpha  \beta  k$ (using the idempotence on $\mathbb{Z}/2\mathbb{Z}$ implying $\alpha^2 =\alpha$ and $\beta^2=\beta$). For each gate, identifying coefficients of $P_{\alpha\beta}(0)$ with those of $1-\mu(\alpha,\beta)$ yields the set of structural parameters as presented in the fourth column of Table \ref{gates}. The fifth column of the table gives the corresponding characteristic polynomial. 

Note that for the $\mu(\alpha,\beta)$ expression for XOR, this identification gives $e=k$ and $e^2\neq k^2$, so that no solution exists with $P_{\alpha\beta}(0) \propto 1-\mu(\alpha,\beta)$ issued from the calculating block \eqref{eq:hamiltonianH0}. But as demonstrated in \cite{dridi2}, other $3\times 3$ calculating blocks can be found for the XOR and one of the corresponding polynomials is given in Table I last column. Indeed, one can find families of polynomials in $\alpha$ and $\beta$ not related to $\mu(\alpha,\beta)$ which take the value 0 if and only if the output is 1, and then determine structural parameters for matrices such that the determinant coincides with these polynomials. This is presented also for the AND, OR and the NOR in the last column of Table I. As presented in Sec \ref{generalization}, increasing the matrix size opens the access to more general non-Boolean polynomial families for each 2-inputs/1-output logic gates, for example the AND in \eqref{eq:Hand}--\eqref{detAnd4} and the XOR gate in \eqref{eq:Hxor}--\eqref{detXor4}.

\subsection{The half adder at a fixed reading energy}
To go beyond the above QHC elementary gates, let us now consider the more involved case of the half adder. It is a two-input and two-output gate. In the discrete ring $\mathbb{Z}/2\mathbb{Z}\equiv \{0,1\}$, the two outputs, sum and carry, are respectively given by $S(\alpha,\beta)=\alpha +\beta-2\alpha \beta$ (which corresponds to XOR) and by $C(\alpha,\beta)= \alpha\beta$ (which corresponds to AND)(see Table \ref{gates}).

Following the fixed-energy output reading approach, the minimum number of states in the calculating block to design a QHC Boolean half adder was found to be 4 \cite{dridi1}. However, this minimal design has the drawback that each reading block needs to be coupled to two states of the calculating block, which is not practical for realistic implementations \cite{dridi1}. We show below that a $5\times 5$ calculating block can be constructed for the half adder, in which the reading block of each logical output is coupled with only one state of this calculating block.

Fixing the reading energy at $E=0$ for simplicity, it was demonstrated above that AND and XOR can both be constructed using $3\times 3$ calculating blocks. However it is not possible to merge those two $3\times 3$  $H_0$ to get a calculating block for the QHC half adder. We therefore define $4\times 4$ calculating blocks for the AND and XOR, namely

\begin{equation} 
\label{eq:Hand}
H_0(\alpha,\beta)_{\text{AND}}= \left[ 
    \begin{matrix}
      0 & 1 & \alpha & -x \\
      1 & 0 & \beta   & -x \\
       \alpha & \beta &-1&x \\
       -x & -x & x & -x^2
      \end{matrix}
    \right],
\end{equation}
whose determinant is given by 
\begin{equation}
\label{detAnd4}
\det[H_0(\alpha,\beta)_{\text{AND}}]=x^2[-2+3(\alpha+\beta)-4\alpha \beta)], 
\end{equation}
and 
\begin{equation} \label{eq:Hxor}
   H_0(\alpha,\beta)_{\text{XOR}}=  \left[ 
    \begin{matrix}
       0 & 1 & \alpha & -x \\
      1 & 0 & \beta   & -x \\
       \alpha & \beta &-1&0 \\
       -x & -x & 0 & x^2      \end{matrix}
    \right],
\end{equation}


\begin{figure}[!t]
    \centering
    \includegraphics[width=.8\linewidth]{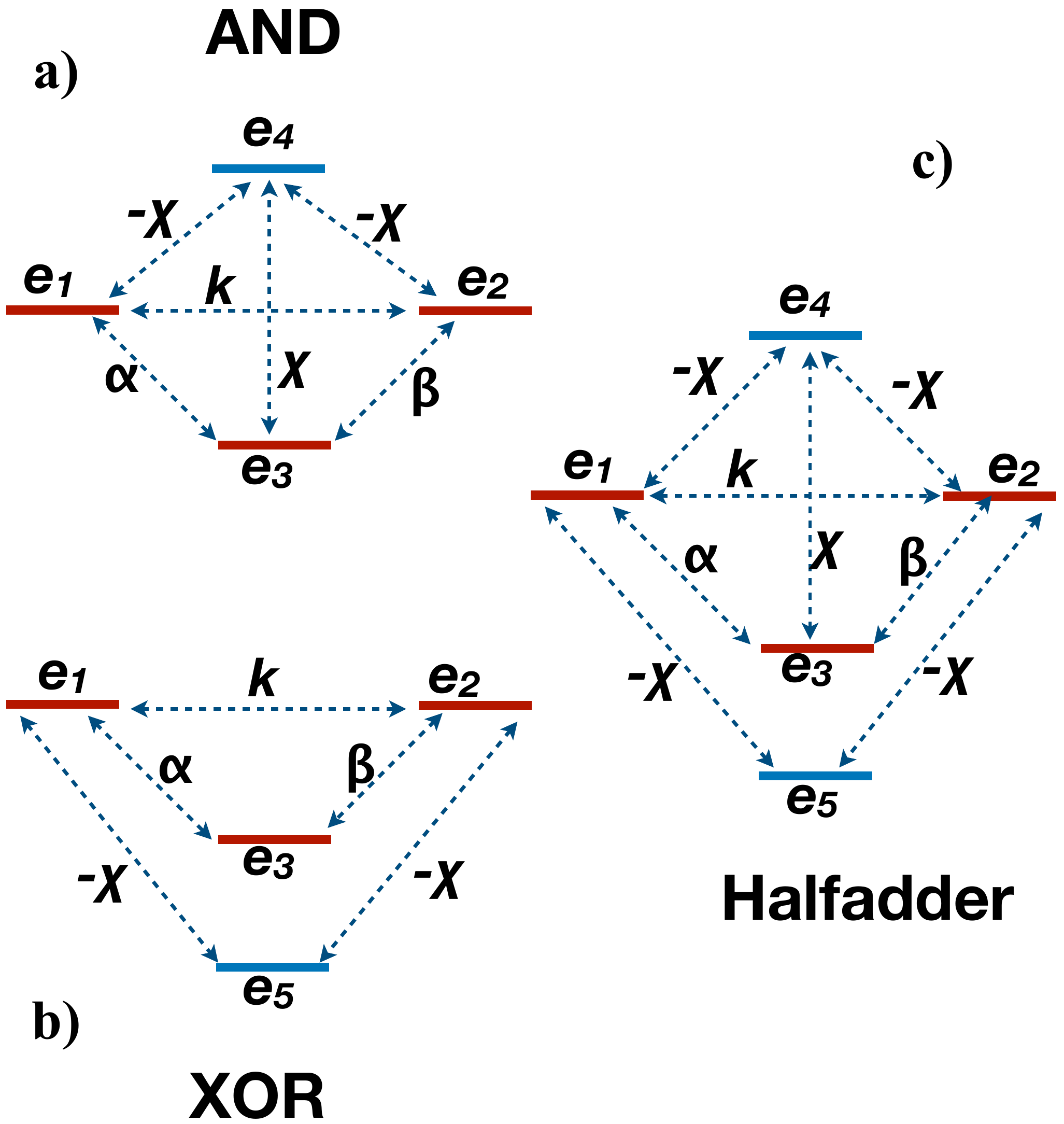} 
   \caption{The quantum graph representations of a) the AND \eqref{eq:Hand}, b) the XOR \eqref{eq:Hxor} and c) the half adder \eqref{eq:Hhalfadder}, the latter resulting from the merging of the AND and XOR graphs.}
  \label{fig:merging}
  \end{figure}

whose determinant is given by 
\begin{equation}
\label{detXor4}
\det[H_0(\alpha,\beta)_{\text{XOR}}]=x^2(-1+\alpha^2+\beta^2),
\end{equation}
with $x$ a free parameter. The quantum graphs of those $4\times 4$ elementary $H_0$ are given in Fig.~\ref{fig:merging}. A way of finding the matrices \eqref{eq:Hand} and \eqref{eq:Hxor} is to first fix the expression of the polynomials \eqref{detAnd4} and \eqref{detXor4}, which take the value 0 if and only if the output is 1, and then to find the structural parameters in the Hamiltonian such that its determinant coincides with these polynomials. 
Note that the determinants \eqref{detAnd4} and \eqref{detXor4} do not follow the condition $P_{\alpha\beta}(E)=0$ for output 1. They belong to the non-Boolean class of polynomials for a $4\times 4$ $H_0(\alpha,\beta)$ calculating block. Of course, the product of their determinants does follow the condition, since 
\begin{equation}
\det[H_0(\alpha,\beta)_{\text{AND}}] \det[H_0(\alpha,\beta)_{\text{XOR}}]=2x^4(1-\alpha)(1-\beta).
\end{equation}

This is a direct consequence of the truth table of the half adder, together with the fact that in $\mathbb{Z}/2\mathbb{Z}\equiv \{0,1\}$, we can use the idempotence relations $\alpha^n =\alpha$ and $\beta^n=\beta$ for the two inputs.
 \begin{figure}[!b]
    \centering
    \includegraphics[width=\linewidth]{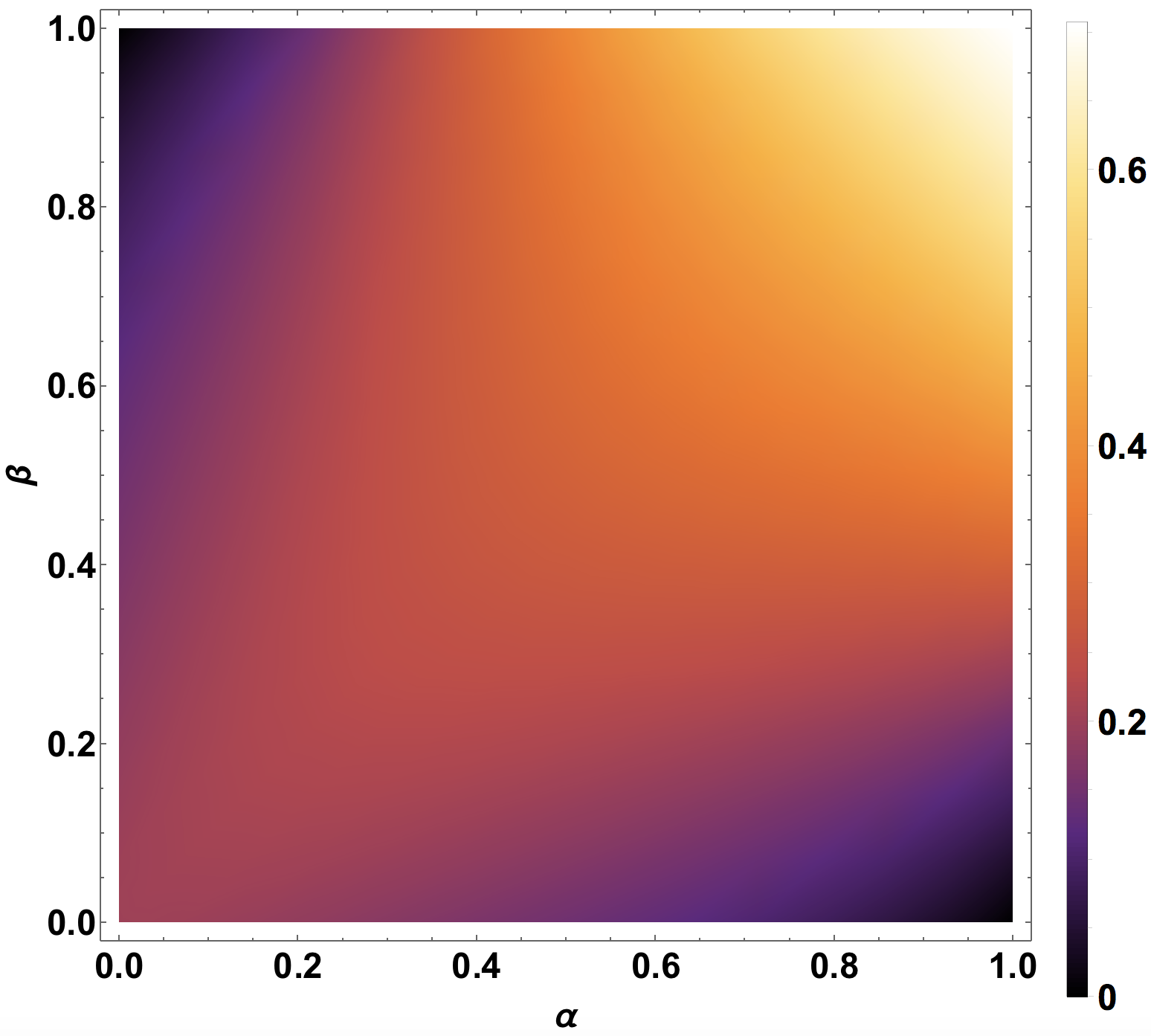}
    \caption{The AND logic response (maximum normalized weight on state $\ket{4}$ in time) as a function of inputs $\alpha , \beta$ calculated from kernel of the $5 \times 5$ half adder Hamiltonian in Eq.~(\ref{eq:Hhalfadder}) at $E=0$ with chosen structural parameter $x=1$.}
    \label{fig:ANDresponse}
  \hspace{0.5cm}
    \centering
    \includegraphics[width=\linewidth]{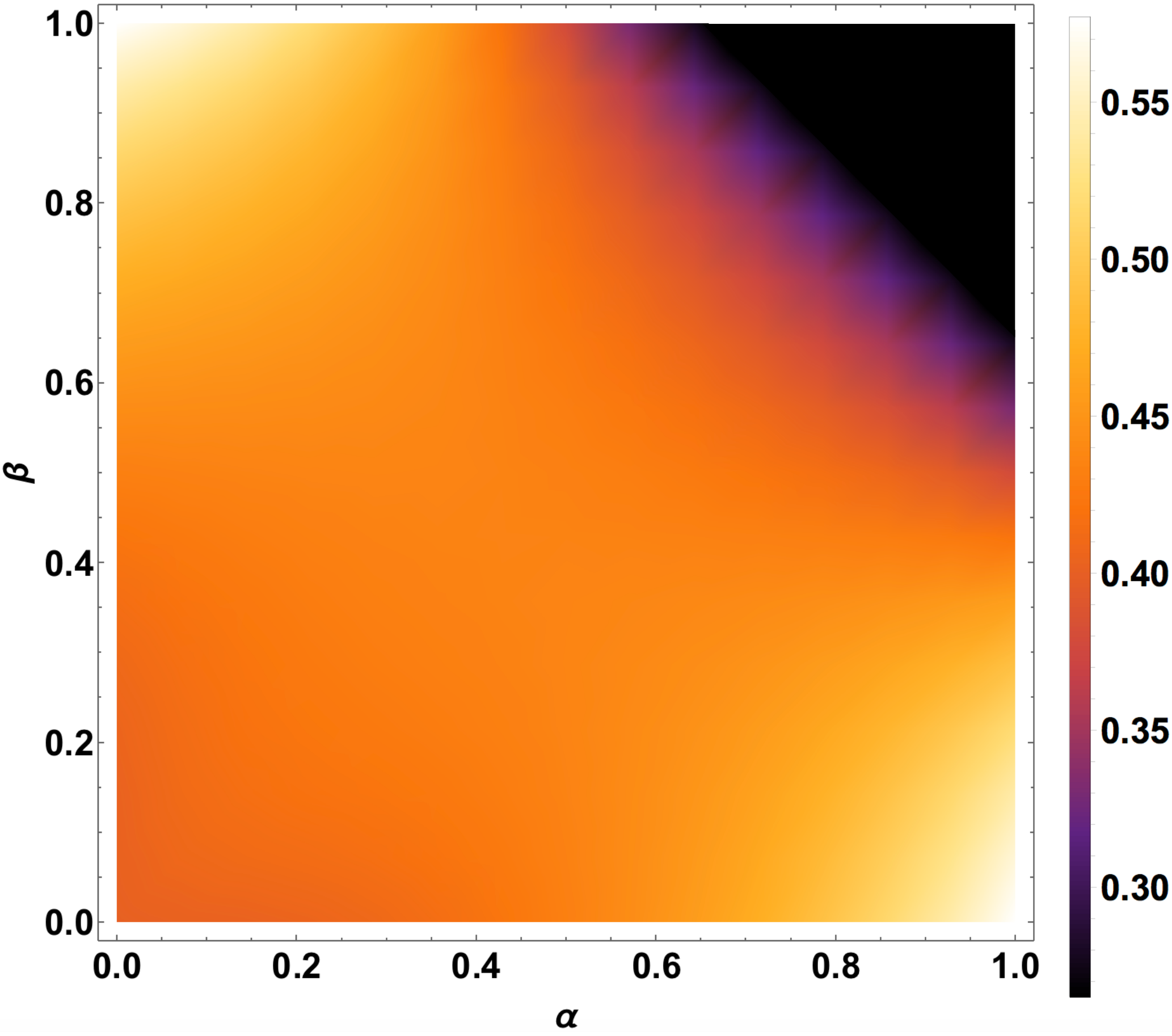}
    \caption{The XOR logic response (maximum normalized weight on state $\ket{5}$ in time) as a function of inputs $\alpha , \beta$ calculated from kernel of the $5 \times 5$ half adder Hamiltonian in Eq.~(\ref{eq:Hhalfadder}) at $E=0$ with chosen structural parameter $x=1$.}
    \label{fig:XORresponse}
\end{figure}


The rationale behind the choice of $4\times 4$ matrices rather than $3\times 3$ is that these two $H_0$ now have the same $3\times 3$ upper-left block. In order to build a calculating block for the half adder, we can merge these 2 elementary $H_0$ into a new $5\times 5$ calculating block:

\begin{equation} 
\label{eq:Hhalfadder}
H_0(\alpha,\beta)_{\text{HA}} 
=\left[ 
 \begin{matrix}
  0 & 1 & \alpha & -x &-x \\
  1 & 0 & \beta   & -x &-x \\
  \alpha & \beta &-1&x &0\\
  -x & -x & x & -x^2 &0\\
  -x & -x&0&0&x^2
  \end{matrix}
\right],
\end{equation}

corresponding to graph c in Fig.~\ref{fig:merging}. 
  
This merging leading to the $5\times 5$ calculating block \eqref{eq:Hhalfadder} for the half adder QHC Hamiltonian can be expressed in mathematical terms by

\begin{equation} 
\label{eq:HandHxor}
H(\alpha,\beta)_{\text{HA}} =VH_0(\alpha,\beta)_{\text{AND}}V^T+WH_0(\alpha,\beta)_{\text{XOR}}W^T-\mathcal{I}(\alpha,\beta).
\end{equation}

Here we introduced
\begin{equation} 
V=\left(
\begin{array}{cccc}
 1 & 0 & 0 & 0 \\
 0 & 1 & 0 & 0 \\
 0 & 0 & 1 & 0 \\
 0 & 0 & 0 & 1 \\
 0 & 0 & 0 & 0 \\
\end{array}
\right),\quad W=\left(
\begin{array}{cccc}
 1 & 0 & 0 & 0 \\
 0 & 1 & 0 & 0 \\
 0 & 0 & 1 & 0 \\
 0 & 0 & 0 & 0 \\
 0 & 0 & 0 & 1 \\
\end{array}
\right),
 \end{equation}
 
 \begin{figure}[!b]
    \centering
    \includegraphics[width=.5\linewidth]{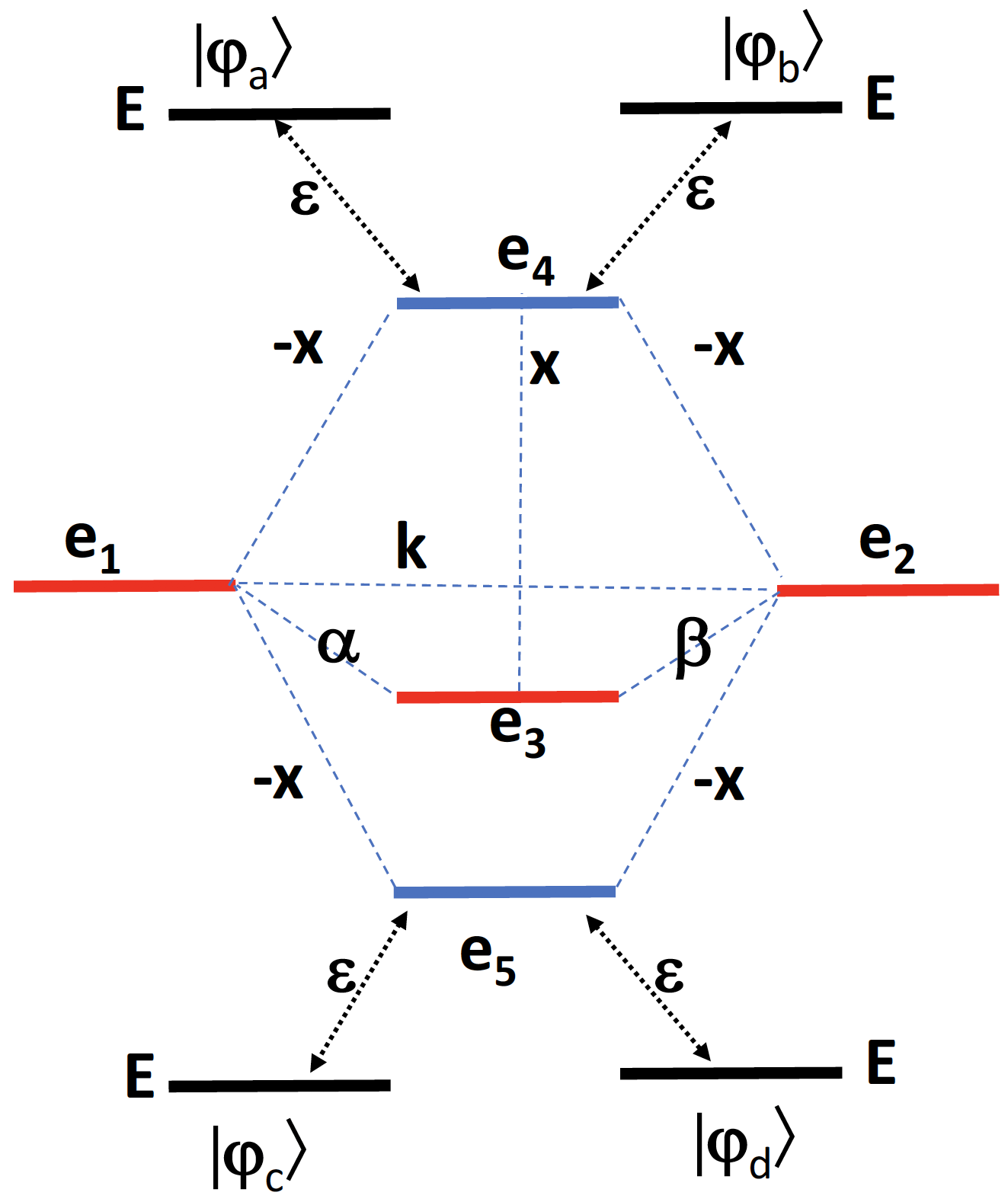} 
   \caption{The QHC quantum graph of the $H(\alpha,\beta)$ Hamiltonian on its complete canonical basis set $ \ket{\varphi_a}, \ket{\varphi_b}, \ket{1}, \ket{2},\ket{3},\ket{4}, \ket{5}, \ket{\varphi_c} ,\ket{\varphi_d}$ with its calculating block defined on $\{ \ket{1}, \ket{2},\ket{3},\ket{4}, \ket{5} \}$ and two reading blocks on $\{ \ket{\varphi_a}, \ket{\varphi_b} \}$ and $\{ \ket{\varphi_c}, \ket{\varphi_d} \}$, for AND and XOR respectively.  $(\alpha,\beta)$ are the classical logical inputs of this QHC half adder. The coupling $k$ and the energy $e$ are the structural parameters used to obtain the Boolean logical functioning for the output reading energy $E = 0$ for both $\{ \ket{\varphi_a}, \ket{\varphi_b} \}$ and $\{ \ket{\varphi_c}, \ket{\varphi_d}\}$.}
  \label{preparation_graphe}
  \end{figure}
  
  
 which increase the size of matrices from $4\times 4$ to $5\times 5$, and 
\begin{equation} 
\mathcal{I}(\alpha,\beta)=\left(
\begin{array}{cccc}
 0 & 1 & \alpha & 0 \\
 1 & 0 & \beta & 0 \\
 \alpha & \beta & -1 & 0 \\
 0 & 0 & 0 & 0 \\
 0 & 0 & 0 & 0 \\
\end{array}
\right),
\end{equation}

which avoids counting twice the inputs.

Fig.~\ref{fig:merging} shows a schematic representation of \eqref{eq:HandHxor}. Three states with red colors with energies $e_1 , e_2$ and $e_3$ are shared. The two reading block have to be connected at $e_5$ for XOR and $e_4$ for AND respectively. 

Similarly to the elementary gates, there is a correspondence on the $\mathbb{Z}/2\mathbb{Z}\equiv \{0,1\}$ ring between the characteristic polynomial of $H_0(\alpha,\beta)_{\text{HA}}$ and the Boolean algebra expression for the outputs. Indeed, the characteristic polynomial must be zero whenever one of the outputs is equal to 1. At energy $E=0$ for the reading blocks, one should have $\det[H_0(\alpha,\beta)_{\text{HA}}]=P_{\alpha\beta}(0)\propto (1-S)(1-C)$, with $S$ and $C$ the sum and carry outputs. But from Table \ref{gates} it comes $S=\alpha +\beta-2\alpha \beta$ and $C= \alpha\beta$ leading directly to $(1-S)(1-C)=(1-\alpha)(1-\beta)$, while $\det[H_0(\alpha,\beta)_{\text{HA}}] =2x^4(1-\alpha)(1-\beta)$. Thus the $H_0(\alpha,\beta)_{\text{HA}}$ determinant is indeed proportional to $(1-S)(1-C)$.

 \begin{figure}[!t]
    \centering
    \includegraphics[width=\linewidth]{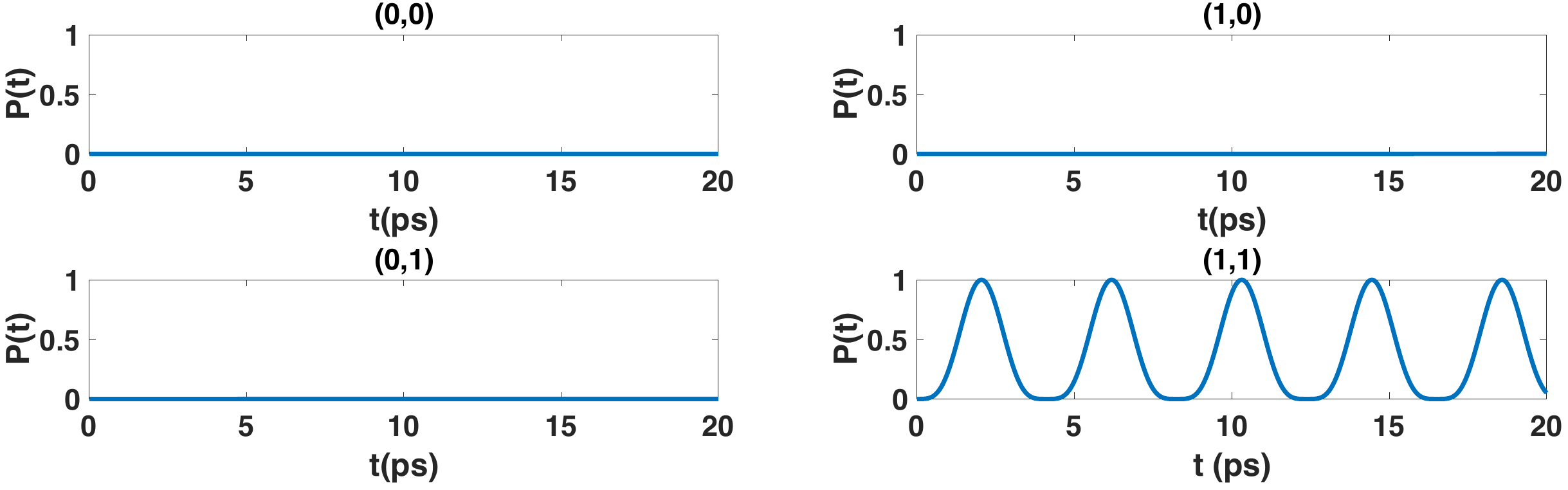} 
   \caption{Time-dependent variations of the populations between $\{ \ket{\varphi_a}, \ket{\varphi_b} \}$ states as a function of the logical input
configurations $(\alpha,\beta)$ (indicated on each curve) after preparing the quantum system  in the non-stationary initial state $ \ket{\varphi_a}$
for the AND. The values of the electronic couplings are $\varepsilon=10^{-3} , x=1$ eV and the
total measuring time is $20 \, \textit{ps}$ .  Note that for t > $4 \, \textit{ns}$ the population for the $(0,0)$ input and for t > $2 \, \textit{ns}$ the population for the $(0,1); (1,0)$ inputs also start to oscillate, reaching the amplitude unity at longer times, confirming that QHC can be based on a frequency and not on a population quantum control. }
  \label{fig:HAab}
  \end{figure}
  
   \begin{figure}[!t]
    \centering
    \includegraphics[width=\linewidth]{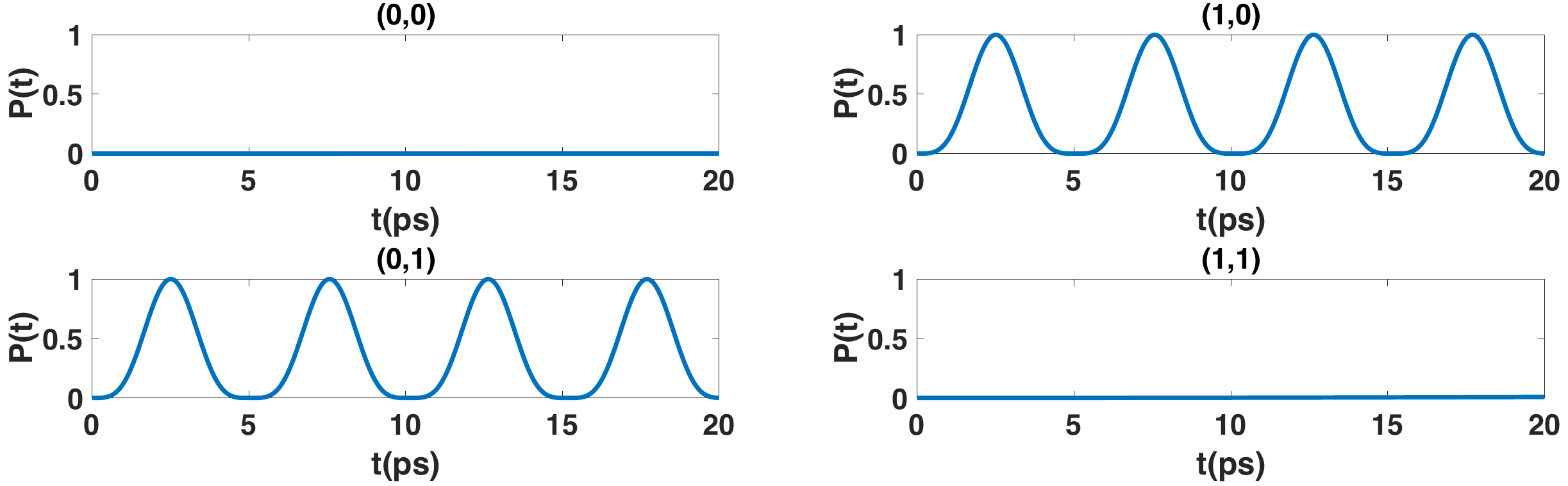} 
   \caption{Time-dependent variations of the populations between $\{ \ket{\varphi_c}, \ket{\varphi_d} \}$ states as a function of the logical input
configurations $(\alpha,\beta)$ (indicated on each curve) after preparing the quantum system  in the non-stationary initial state $\ket{\varphi_c}$
for the XOR. The values of the electronic couplings are $\varepsilon=10^{-3} , x=1$ eV and the
total measuring time is $20 \, \textit{ps}$ . Note that for t > $2 \, \textit{ns}$ the population for the $(0,0)$ input and for t > $0.66  \, \textit{ns}$ the population for the $(1,1)$ input also start to oscillate, reaching amplitude unity at longer times.}
  \label{fig:HAcd}
  \end{figure}
  

In order to check that $H_0(\alpha,\beta)_{\text{HA}}$ yields a half adder, we also have to check that for input $(0,0)$ the kernel is empty (which is the case), while for inputs $(0,1)$, $(1,0)$ and $(1,1)$ it must have the correct weight on the output states $\ket{4}$ (with energy $e_4$) and $\ket{5}$ (with energy $e_5$). For these inputs, the kernel is one-dimensional and generated by the eigenvectors $V(\alpha,\beta)_{\text{HA}}$ given by
\begin{equation} 
 \begin{aligned}
V(0,1)_{\text{HA}}&=\frac{1}{\sqrt{1+2x^2}}\{0,x,x,0,1\}\\
V(1,0)_{\text{HA}}&=\frac{1}{\sqrt{1+2x^2}}\{x,0,x,0,1\}\\
V(1,1)_{\text{HA}}&=\frac{1}{\sqrt{1+x^2}}\{0,0,x,1,0\}.
 \end{aligned}
\end{equation}
For inputs $(0,1)$ and $(1,0)$ there is zero weight on $\ket{4}$ and a nonzero weight on $\ket{5}$, while for input $(1,1)$ it is the reverse, which is the desired property. Thus $H_0(\alpha,\beta)_{\text{HA}}$ is indeed the calculating block of a half adder. In order to account for possible imperfections in the experimental implementation, we show in Fig.~\ref{fig:ANDresponse} how the weight on $\ket{4}$ depends on the value of inputs $\alpha$ and $\beta$, and similarly for $\ket{5}$ in Fig.~\ref{fig:XORresponse}. It shows that there is a significant area of parameters where the half adder is working properly.

One can also directly check that our half adder gate works properly by constructing the full Hamiltonian of the QHC molecule by adding the two reading blocks of size $2\times 2$, leading to the final $9\times 9$ QHC Hamiltonian of the half adder whose quantum graph is presented in Fig.~\ref{preparation_graphe}. The time-dependent Heisenberg-Rabi oscillations of each reading block are presented in Figs.~\ref{fig:HAab} and \ref{fig:HAcd}.

\section{Generalization of the fixed energy approach}
\label{generalization}
\subsection{The basic principles}
In order to go beyond the $H_0(\alpha,\beta)$ matrices constructed in the previous sections, a more systematic approach can be followed. Indeed, these matrices have the peculiar block structure
\begin{equation}
\label{matrixH}
H_0(\alpha,\beta)=\left(
\begin{array}{c|c}
 \begin{array}{ccc}&&\\ &C_{\alpha\beta}& \\&&\end{array} &  \begin{array}{c} \phantom{A}\\ A\\ \phantom{A}\end{array} \\
 \hline
 \begin{array}{ccc}&A^T&\end{array} &  \begin{array}{c}B \end{array} 
 \end{array}
\right),
\end{equation}
where we have separated the quantum states in the QHC graph which depend on the input (described by the matrix $C_{\alpha\beta}$) and the ones through which the reading is performed (which are associated with matrix $B$). For example for the AND gate \eqref{eq:Hand} having a single reading state, $A^T$ is a $1\times 3$ matrix, i.e.~a single vector $u=(-x,-x,x)$. For the XOR gate, \eqref{eq:Hxor}, $A^T$ is the vector $v=(-x,-x,0)$. For the two-output half adder, $A$ is a two-column matrix whose columns are $u$ and $v$, and $B$ is a $2\times2$ matrix. 

One can therefore fix the matrix $C_{\alpha\beta}$ and find the equations that $A$ and $B$ have to satisfy in order that $H_0$ produce the expected Boolean logic gate:
\begin{itemize}
\item if the output is 1, an eigenvalue $E=0$ with an eigenvector having a nonzero component on the reading state, 
\item if the output is 0, either an eigenvalue $E=0$ with eigenvector having a zero component on the reading state or no eigenvalue equal to 0. 
\end{itemize}
Moreover and if it exists, the $H_0(\alpha,\beta)$ kernel must be one-dimensional (otherwise a linear combination of its vectors may allow to modify arbitrarily the component on the reading state).

\subsection{AND gate}
In order to illustrate this approach, let us recover the result of \eqref{eq:Hand} for the AND gate. First, the $H_0$ upper-left block in \eqref{matrixH} is given by
 \begin{equation}
\label{matrixCand}
C_{\alpha\beta}=\left(
 \begin{array}{ccc}0&1&\alpha\\ 1&0&\beta\\ \alpha&\beta&-1
 \end{array}
\right),
\end{equation}
which is invertible for any of the four possible inputs. Matrix $A$ is a column vector $u^T$ which must be determined. For input $(\alpha,\beta)=(1,1)$ the output of AND is 1, thus the kernel of $H_0(1,1)$ should be one-dimensional and contain a vector of the form $z=(p_1,p_2,p_3,1)$ so that it has a non-zero component (which we fix equal to 1) on the reading block. From \eqref{matrixH}, the condition that $z$ belongs to the kernel of $H_0(1,1)$ is equivalent to $C_{11}p+ u=0$ and $u^Tp+B=0$ (in this case $B$ is just a number), with $p=(p_1,p_2,p_3)$. The first condition gives $p=-C_{11}^{-1}u$. The second equation then gives $B=u^TC_{11}^{-1}u$. Therefore any nonzero vector $u$ yields a matrix with the suitable property, provided for the other inputs the kernel is empty (which generically is true). For instance, choosing $u=(-x,-x,x)$ yields $B=- x^2$, which corresponds to the solution found in the last section. Choosing $u=(x,x,x/2)$ yields $B=5x^2/4$, and the corresponding polynomial $P_{\alpha\beta}(0)=-\frac12 x^2 (1-\alpha\beta)$ is proportional to $1-\mu(\alpha,\beta)$ (so that the matrix given by this solution is the ''Boolean algebra'' one). More generally, we have now a 3-parameter family of solutions, parametrized by the entries of $u$. Namely, if we denote $u=(u_1,u_2,u_3)$ we have $B=\frac{1}{3} \left(-u_1^2+2 \left(2 u_2+u_3\right) u_1-\left(u_2-u_3\right){}^2\right)$, and the determinant reads
\begin{align}
\label{polyANDgeneral}
P_{\alpha\beta}(0)&=\alpha  u_2
   \left(u_2-2 u_3\right)+\beta  u_1 \left(u_1-2 u_3\right)\nonumber\\
&-\frac{2}{3} \alpha  \beta  \left(u_1^2-\left(u_2+2 u_3\right) u_1+\left(u_2-u_3\right){}^2\right)\nonumber\\
&+\frac{1}{3} \left(-u_1^2-2 \left(u_2-u_3\right)
   u_1-u_2^2+2 u_3^2+2 u_2 u_3\right).
\end{align}
For any given polynomial in $\alpha$ and $\beta$ taking the value 0 if and only if the output is 1, one just needs to look for values of the $u_i$ such that the determinant \eqref{polyANDgeneral} coincides with this polynomial. This leads to a new non-Boolean $P_{\alpha\beta}(0)$ example as in the last column of Table 1 but here constructed from a $4\times 4$ $H_0(\alpha,\beta)$ calculating block. 

\subsection{Half adder}
Let us now consider the case of the half adder, which has two outputs, sum and carry, or equivalently an output for AND and one for XOR. We denote the output string by $(\mu,\nu)$. Section \ref{booleanalgebra} yields a solution, \eqref{eq:HandHxor}, with $C_{\alpha\beta}$ given by the $3\times 3$ matrix \eqref{matrixCand}. In order to find the most general solution, we may consider that $H_0$ is an $N\times N$ Hamiltonian with block structure \eqref{matrixH}, in which $C_{\alpha\beta}$ is now a fixed $(N-2)\times (N-2)$ block. 
Since for the three inputs $(\alpha,\beta)=(0,1)$, $(1,0)$ and $(1,1)$ at least one of the outputs is 1, the kernel of $H_0$ has to be non-empty (and one-dimensional) for these three inputs. Moreover the corresponding vector must have a nonzero entry whenever the output is 1 and a 0 entry whenever it is 0 : for output $\mu$ we can simply fix that entry to the value $\mu$. To account for conditions for both outputs the vector $z$ in the kernel of $H_0$ must therefore be of the form $z=(p_1,p_2,\ldots,p_{N-2},\mu,\nu)$ for an input $(\alpha,\beta)$ and output $(\mu,\nu)$. The condition that $z$ belongs to the kernel can be rewritten
\begin{eqnarray}
\label{cp}
C_{\alpha\beta}p+\mu u+\nu v&=&0\\
\label{up}
u p+\mu B_{11}+\nu B_{12}&=&0\\
\label{vp}
v p+\mu B_{21}+\nu B_{22}&=&0,
\end{eqnarray}
with $p=(p_1,\ldots,p_{N-2})$ (from now on, for simplicity of notations, we will remove the distinction between $u$ and $u^T$, since which is which is obvious from the context). For simplicity we will again assume that the matrix $C_{\alpha\beta}$ is invertible for all inputs. Then \eqref{cp} gives a unique solution $p=-C_{\alpha\beta}^{-1}(\mu u+\nu v)$. Inserting this solution into \eqref{up}--\eqref{vp} gives
\begin{eqnarray}
\label{up2}
\mu (B_{11}-uC_{\alpha\beta}^{-1}u)+\nu (B_{12}-uC_{\alpha\beta}^{-1}v)&=&0\\
\label{vp2}
\mu (B_{21}-vC_{\alpha\beta}^{-1}u)+\nu (B_{22}-vC_{\alpha\beta}^{-1}v)&=&0.
\end{eqnarray}
In  \eqref{up2}--\eqref{vp2} we recognize entries of the Schur complement \cite{Gollub} of block  $C_{\alpha\beta}$ in matrix $H_0$, defined by $S_{\alpha\beta}=B-AC_{\alpha\beta}^{-1}A^T$ (we recall that $u$ and $v$ are the column vectors of $A$). In terms of the Schur complement, \eqref{up2}--\eqref{vp2} can be reformulated as
\begin{equation}
\label{defS}
S_{\alpha\beta}\left(
\begin{array}{c}\mu\\ \nu\end{array}
\right)=0,
\end{equation}
that is, the vector $(\mu,\nu)$ of outputs belongs to the kernel of $S_{\alpha\beta}$. We can now examine what condition \eqref{defS} implies for each input. For input $(1,1)$, we have that output $(1,0)\in $Ker$S_{11}$, i.e.~the first column of $S_{11}$ is zero, leading to conditions
\begin{equation}
\label{output01h}
(B-AC_{11}^{-1}A^T)_{11}=0,\quad (B-AC_{11}^{-1}A^T)_{21}=0.
\end{equation}
For inputs $(0,1)$ and $(1,0)$ we have output $(0,1)\in$Ker$S_{01}$ and $(0,1)\in $Ker$S_{10}$, thus the second column of $S_{01}$ and $S_{10}$ must be zero, which means that
\begin{align}
\label{output10ha}
(B-AC_{01}^{-1}A^T)_{12}=0&,\quad(B-AC_{01}^{-1}A^T)_{22}=0,\\
\quad(B-AC_{10}^{-1}A^T)_{12}=0&,\quad(B-AC_{10}^{-1}A^T)_{22}=0.
\label{output10hb}
\end{align}
\eqref{output01h}--\eqref{output10hb} allow to express matrix $B$ (which does not depend on the input) in terms of $u$, $v$ and $C$. They yield
\begin{eqnarray}
\label{matrixBh}
B_{11}=uC_{11}^{-1}u,&\quad B_{12}=uC_{01}^{-1}v=uC_{10}^{-1}v,\nonumber\\
B_{21}=vC_{11}^{-1}u,&\quad B_{22}=vC_{01}^{-1}v=vC_{10}^{-1}v.
\end{eqnarray}
Symmetry of $H_0$, and thus of $B$, implies that $B_{12}=B_{21}$, which yields an additional condition. Thus, for any fixed $A$ and $C$, a matrix $B$ fulfilling conditions \eqref{output01h}--\eqref{output10hb} will exist if and only if
\begin{equation}
\label{condB1h}
uC_{11}^{-1}v=uC_{01}^{-1}v=uC_{10}^{-1}v,\quad vC_{01}^{-1}v=vC_{10}^{-1}v,
\end{equation}
are satisfied. Any pair $(u,v)$ solution of the quadratic polynomial equations \eqref{condB1h} yields a possible half adder Hamiltonian (provided the condition that the kernel is empty for input $(0,0)$ is met). For instance, one can check that $u=(-x,-x,x)$ and $v=(-x,-x,0)$ are solution of \eqref{condB1h}. The corresponding matrix (which is precisely \eqref{eq:Hhalfadder}) is indeed a half adder matrix. As a consequence we can now build whole families of Hamiltonians giving half adders, opening more possibilities for experimental implementations. In the Appendix \ref{app1} we show how to construct more general solutions to these equations.

\subsection{Full adder}
\begin{figure}[!t]
    \centering
    \includegraphics[width=\linewidth]{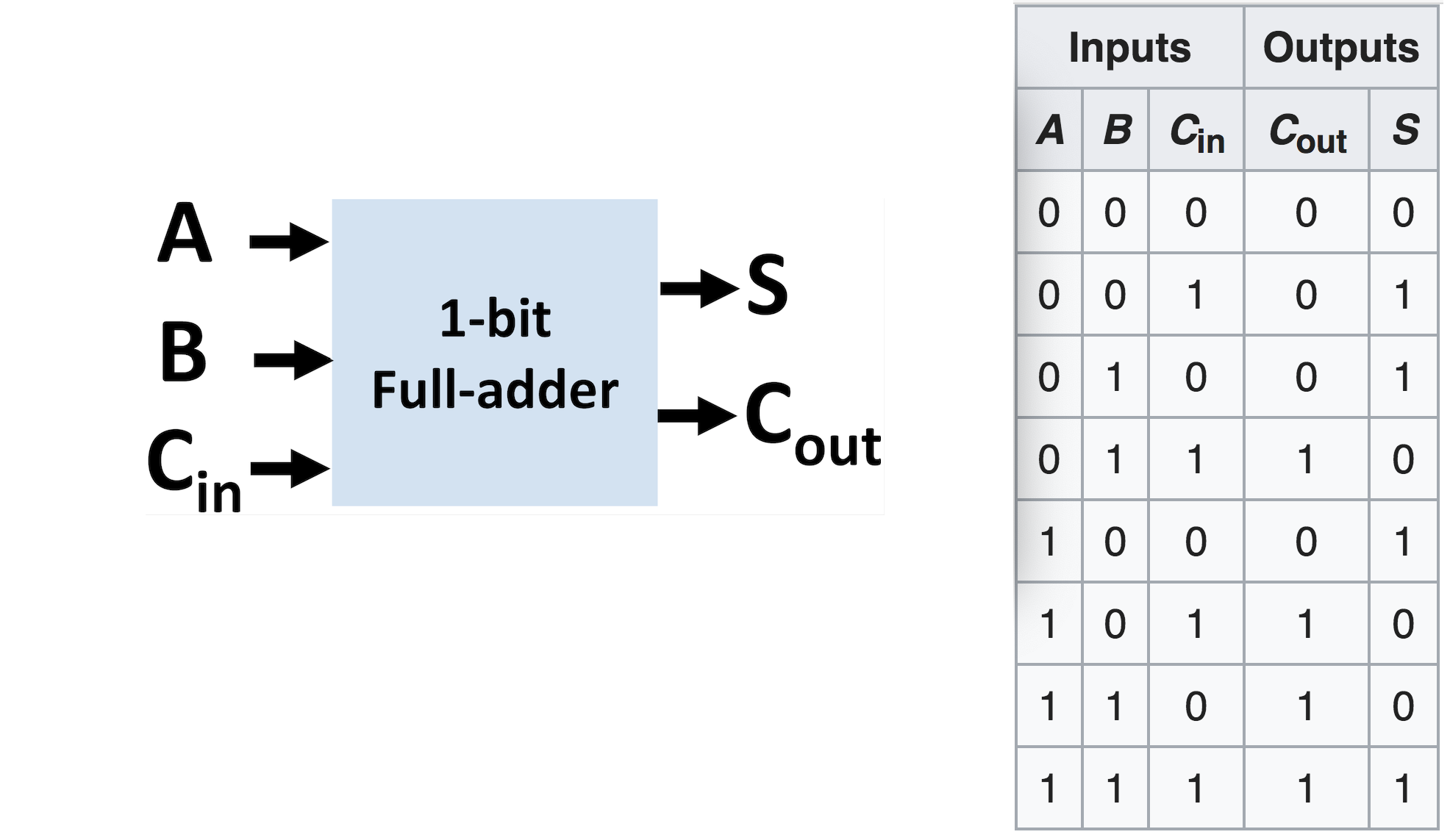} 
   \caption{Schematic drawing for a 1-bit full adder with input carry $C_{\text{in}}$ and output carry $C_{\text{out}}$ and corresponding logical truth table.}
  \label{fig:fulltruth}
  \end{figure}
  
For the elementary Boolean gates of Table \ref{gates}, as well as for the half adder, Hamiltonians $H_0$ were known from direct calculations based on the characteristic polynomial. The generalized approach put forward in the present section has allowed to recover these solutions and to find whole new families. In the case of the full adder, with three inputs $(\alpha,\beta,\gamma)$ and two outputs $(\mu,\nu)$ (sum and carry), no solution was known so far. As we now show, our new method allows to produce families of solutions.

The truth table for the full adder is recalled in Fig.~\ref{fig:fulltruth}. Equations similar to \eqref{matrixH}--\eqref{defS} are easily obtained, where the upper left part of $H_0$ is now $C_{\alpha\beta\gamma}$. They lead to the analog of \eqref{defS},
\begin{equation}
\label{defSfull}
S_{\alpha\beta\gamma}\left(
\begin{array}{c}\mu\\ \nu\end{array}
\right)=0,
\end{equation}
where the $2\times 2$ matrix $S_{\alpha\beta\gamma}=B-AC_{\alpha\beta\gamma}^{-1}A^T$ is the Schur complement of $C_{\alpha\beta\gamma}$. If we consider inputs $(0,0,1)$, $(0,1,1)$ and $(1,1,1)$ in \eqref{defSfull}, we get the conditions that $(1,0)\in$Ker$S_{011}$, $(0,1)\in$Ker$S_{001}$ and $(1,1)\in$Ker$S_{111}$. This translates into the fact that the first column of $S_{011}$ is zero, the second column of $S_{001}$ is zero and the sum of each line of $S_{111}$ is zero. The corresponding equations read
\begin{equation}
\label{output01}
(B-AC_{011}^{-1}A^T)_{11}=0,\quad (B-AC_{011}^{-1}A^T)_{21}=0,
\end{equation} 
\begin{equation}
\label{output10}
(B-AC_{001}^{-1}A^T)_{12}=0,\quad(B-AC_{001}^{-1}A^T)_{22}=0,
\end{equation}
and 
\begin{align}
\label{output11a}
(B-AC_{111}^{-1}A^T)_{11}+(B-AC_{111}^{-1}A^T)_{12}=0,&\\
(B-AC_{111}^{-1}A^T)_{21}+(B-AC_{111}^{-1}A^T)_{22}=0.&
\label{output11b}
 \end{align}
Similar equations are obtained from inputs with permuted strings. The  $2\times 2$ matrix $B$ exists if and only if the symmetry condition $B_{12}=B_{21}$, together with \eqref{output01}--\eqref{output11b} and the ones obtained by permutation of indices, are all compatible with each other. All these compatibility conditions are equivalent to
\begin{align}
\label{rel1}
&u(C_{011}^{-1}-C_{111}^{-1})u+u(C_{001}^{-1}-C_{111}^{-1})v=0\\
&u(C_{011}^{-1}-C_{111}^{-1})v+v(C_{001}^{-1}-C_{111}^{-1})v=0\label{condB2}\\
\label{condt4}
&uC_{011}^{-1}u= uC_{101}^{-1}u=uC_{110}^{-1}u\\
\label{condt5}
&vC_{001}^{-1}v= vC_{010}^{-1}v=vC_{100}^{-1}v\\
&uC_{011}^{-1}v=uC_{101}^{-1}v=uC_{110}^{-1}v=uC_{001}^{-1}v=uC_{010}^{-1}v=uC_{100}^{-1}v,
\label{condt6}
\end{align}
which are again quadratic polynomial equations in $u$ and $v$. One can look for a solution to these equations following the same ideas as for the half adder. Appendix \ref{app2} details the steps that allow to reduce this set of equations to just three equations, which are tractable numerically.

\begin{figure}[!t]
    \centering
    \includegraphics[width=\linewidth]{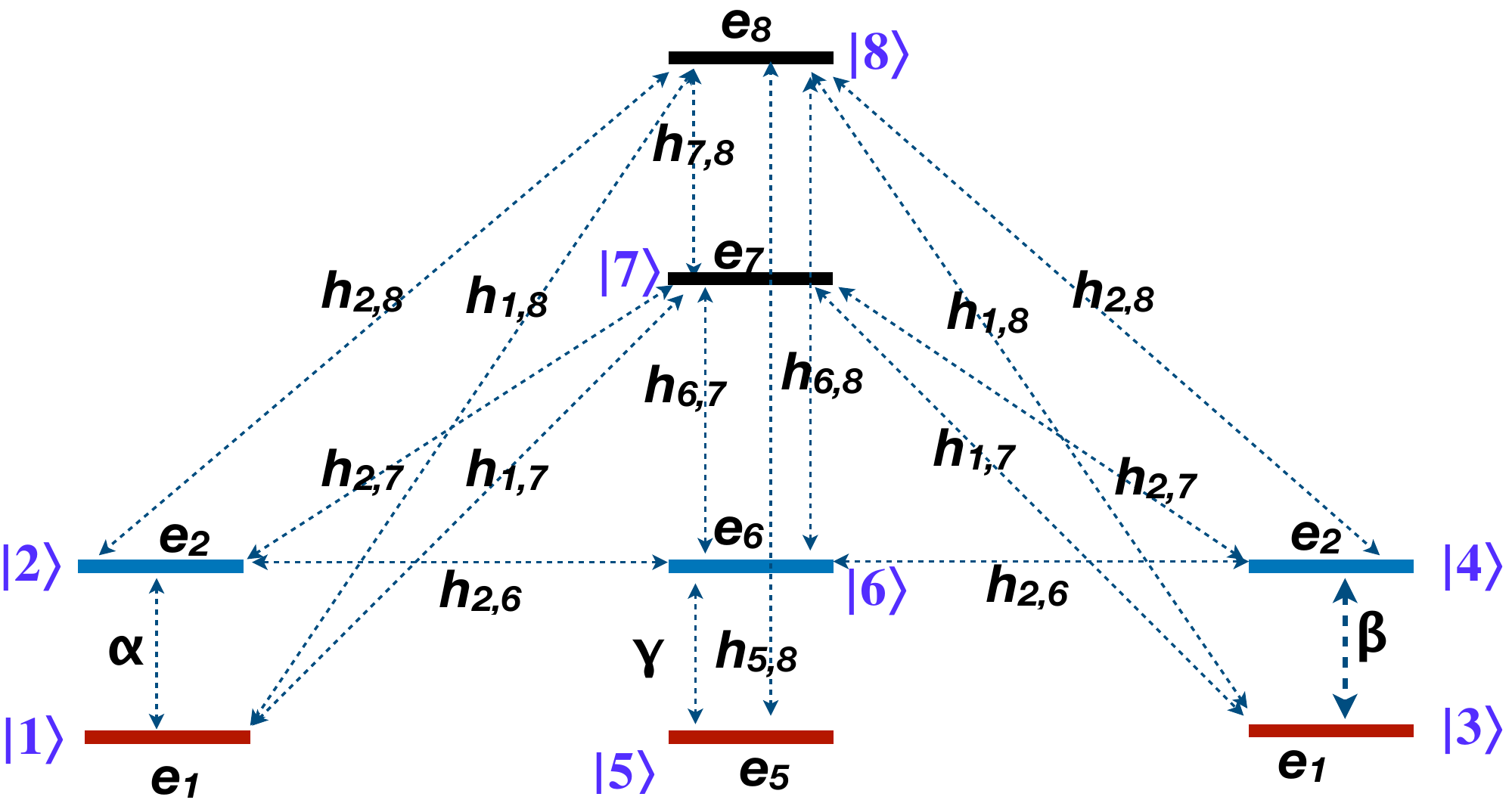} 
   \caption{A $8\times 8$ quantum graph corresponding to the calculating block of a Boolean 3-inputs, 2-outputs full adder in which two reading outputs have to be connected to states 7 and 8 separately.}
  \label{fig:8x8}
  \end{figure}

As an illustration, we look for Hamiltonians $H_0(\alpha,\beta,\gamma)=(h_{i,j})_{1\leq i,j\leq 8}$ with block form \eqref{matrixH} and 
\begin{equation} \label{eq:hamiltonian8x8}
C_{\alpha\beta\gamma} = \left(
\begin{array}{cccccc}
 e_1 & \alpha  & 0 & 0 & 0 & 0  \\
 \alpha  & e_2 & 0 & 0 & 0 & h_{2,6}  \\
 0 & 0 & e_1 & \beta  & 0 & 0  \\
 0 & 0 & \beta  & e_2 & 0 & h_{2,6}  \\
 0 & 0 & 0 & 0 &e_5 & \gamma  \\
 0 & h_{2,6} & 0 & h_{2,6} & \gamma  & e_6 
 \end{array}
\right),
\end{equation}
corresponding to the graph displayed at Fig.~\ref{fig:8x8}.
Applying the above approach, solutions $u$ and $v$ can be found. A typical solution $H^{0}_{\text{typ}}$ reads
\begin{equation} \label{hypfulladder}
\left(
\begin{array}{cccccccc}
 -1 & \alpha  & 0 & 0 & 0 & 0 & -1 & \frac{1}{2} \\
 \alpha  & \frac{1}{2} & 0& 0 & 0& 1 & 0.27 & -1.17 \\
 0 & 0 & -1& \beta  & 0 & 0 & -1 & \frac{1}{2} \\
 0& 0 & \beta  & \frac{1}{2} & 0 & 1 & 0.27 & -1.17 \\
 0& 0 & 0 & 0& -1& \gamma  & 1.56 & 0.66 \\
 0& 1 & 0 & 1 & \gamma  & \frac{1}{2} & -1.40& -2.88 \\
 -1 & 0.27 & -1 & 0.27 & 1.56 & -1.40 & -3.96 & -0.40 \\
 \frac{1}{2} & -1.17 & \frac{1}{2} & -1.17 & 0.66 & -2.88 & -0.40 & 2.12 \\
\end{array}
\right).
\end{equation}

\begin{figure}[!t]
    \centering
    \includegraphics[width=\linewidth]{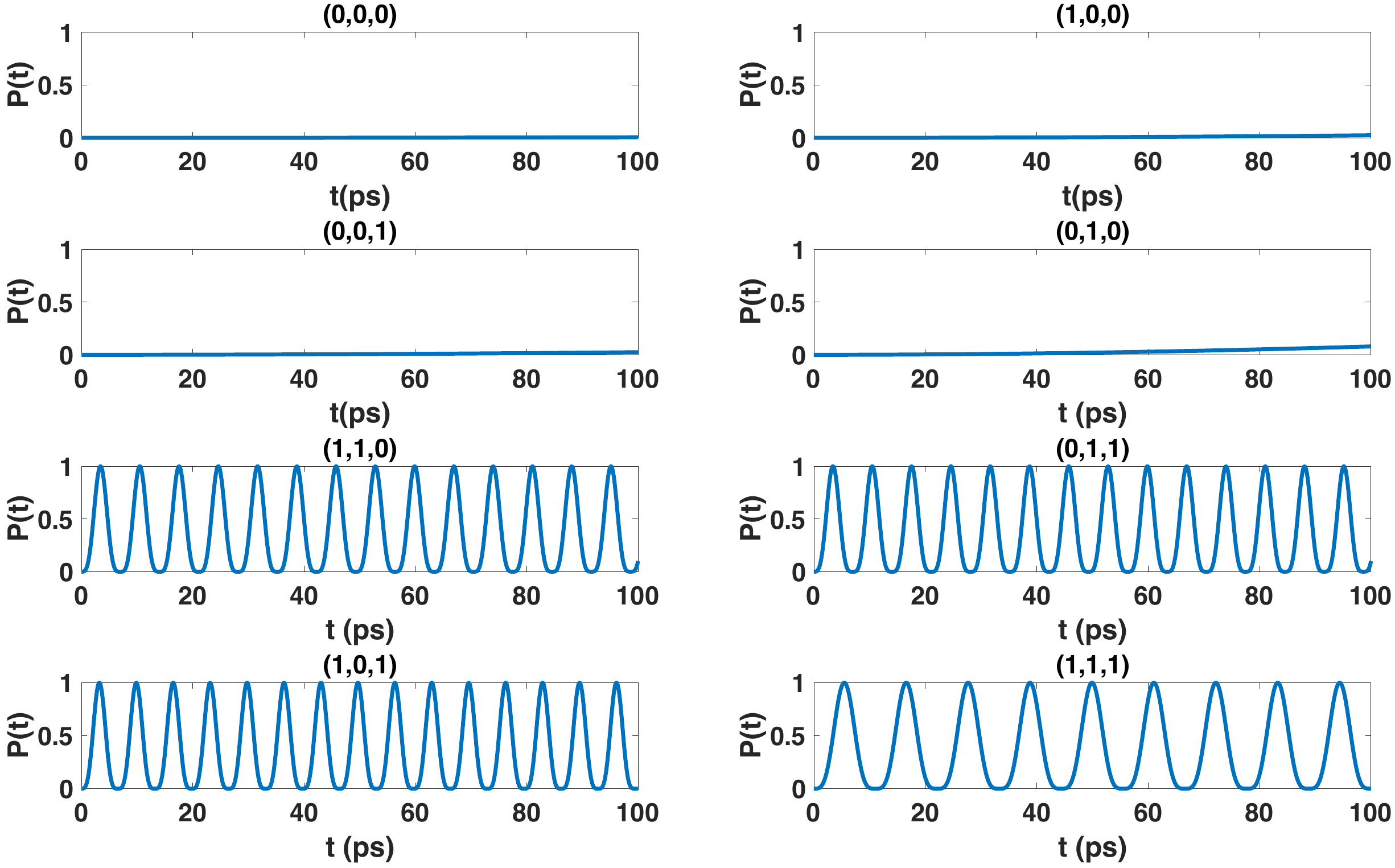} 
   \caption{Time-dependent variations of the $ \ket{\varphi_b}$ population through state $\ket{7}$ with energy $e_7$ (Fig.~\ref{fig:8x8}) as a function of the logical input
configurations $(\alpha,\gamma,\beta)$ (indicated on each curve) after preparing this quantum system in the non-stationary initial state $ \ket{\varphi_a}$ for the $C_{\text{out}}$ output. The values of the electronic couplings are $\varepsilon=10^{-3} $ eV and the total measuring time is $100 \, \textit{ps}$ .}
  \label{fig:fulladder_7}
  \end{figure}

\begin{figure}[!t]
    \centering
    \includegraphics[width=\linewidth]{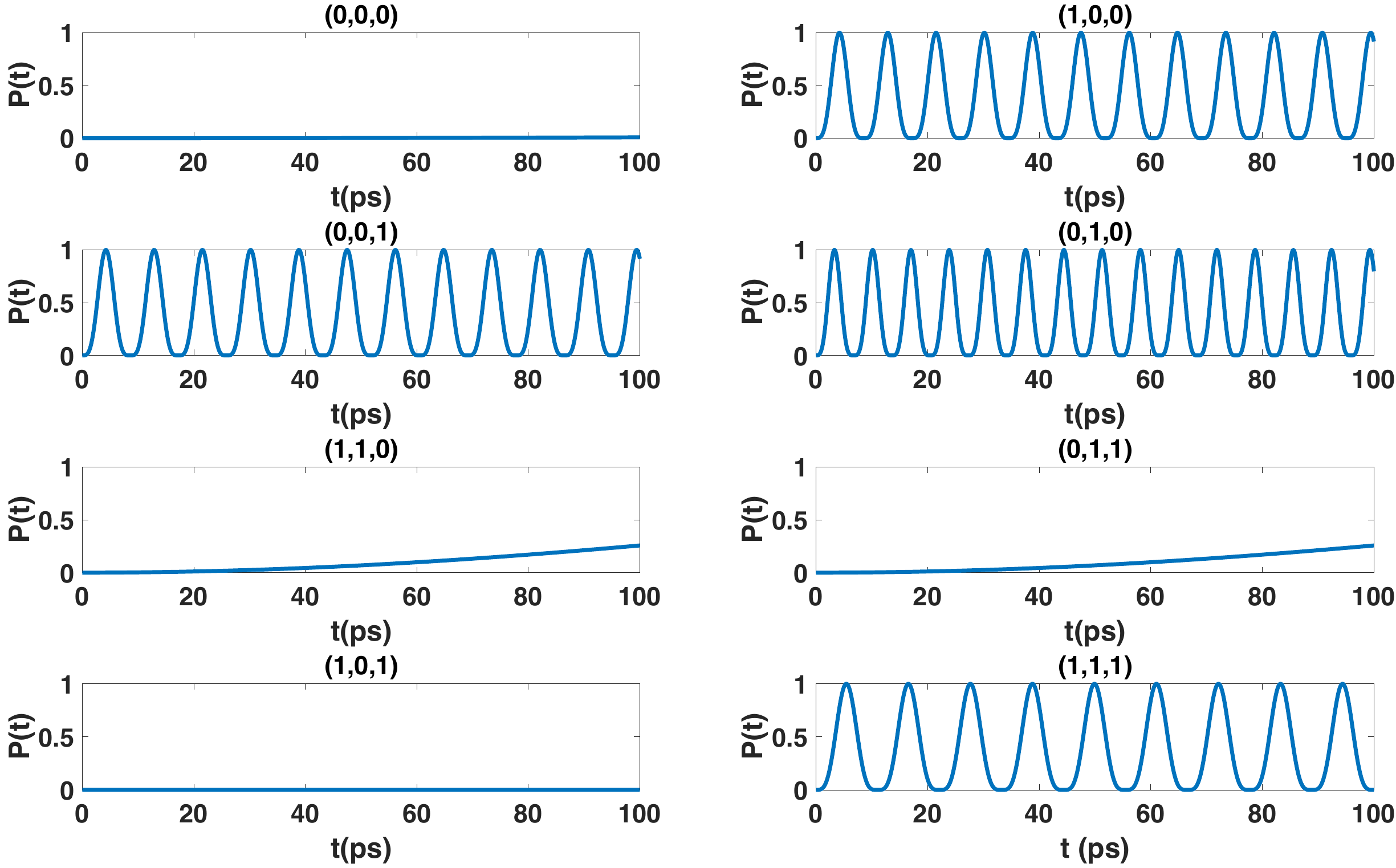} 
   \caption{Time-dependent variations of the $\ket{\varphi_d}$ population through state $\ket{8}$ (Fig.~\ref{fig:8x8}) with energy $e_8$ as a function of the logical input configurations $(\alpha,\gamma,\beta)$ (indicated on each curve) after preparing this quantum system in the non-stationary initial state $ \ket{\varphi_c}$ for the $S$ output. The values of the electronic couplings are $\varepsilon=10^{-3} $ eV and the total measuring time is $100 \, \textit{ps}$.}
  \label{fig:fulladder_8}
  \end{figure}

It remains to be checked that the eigenvectors in the kernel have the correct weight on reading states $\ket{7}$ and $\ket{8}$. There is no zero eigenvalue for input $(0,0,0)$, while for other outputs the kernel is one-dimensional with normalized eigenvectors given by
\begin{equation} \label{evfulladder}
\begin{aligned}
&V_{\text{typ}}^{001}=\{  -0.17, -0.82, -0.33, -0.16, -0.22, 0.01, 0, -0.33\}\\
&V_{\text{typ}}^{100}= \{ -0.33, -0.16, -0.17, -0.82, -0.22, 0.01, 0, -0.33 \}\\
&V_{\text{typ}}^{010}= \{ -0.21, -0.15, -0.21, -0.15, -0.70, -0.42,0, -0.42 \}\\
&V_{\text{typ}}^{011}= \{ 0.41, 0.30, 0.24, -0.17, -0.69, -0.04, -0.41, 0 \}\\
&V_{\text{typ}}^{110}= \{ 0.24, -0.17, 0.41, 0.30, -0.69, -0.04, -0.41, 0 \}\\
&V_{\text{typ}}^{101}= \{ -0.07, 0.36, -0.075, 0.36, 0.69, -0.22, 0.44, 0 \}\\
&V_{\text{typ}}^{111}= \{ -0.07, 0.05, -0.07, 0.05, 0.87, 0.28, 0.26, 0.26 \}
\end{aligned}
\end{equation}
As \eqref{evfulladder} shows, for inputs (0,0,1), (1,0,0) and (1,0,1) there is no weight on $\ket{7}$ and nonzero weight on $\ket{8}$, and vice versa for inputs (0,1,1), (1,1,0) and (1,0,1). For input (1,1,1) there are two identical weights on states $\ket{7}$ and $\ket{8}$. Thus the calculating block $H^{0}_{\text{typ}}$ in \eqref{hypfulladder} achieves the desired truth table. This is the first example of a QHC full adder.

One can check that the determinant of $H^{0}_{{\text{typ}}}$ coincides with the expression obtained from the Boolean algebra analysis. Using again $\alpha^2=\alpha, \beta^2=\beta,\gamma^2=\gamma$, the sum output for the full adder reads
\begin{equation} \label{Sfull}
\begin{aligned}
S(\alpha,\beta,\gamma)&=\alpha +\beta + \gamma +2(2\alpha \beta \gamma -\alpha \beta -\alpha \gamma -\gamma \beta)
\end{aligned}
\end{equation}
and the carry output is
\begin{equation} \label{Cfull}
\begin{aligned}
C_{\text{out}}(\alpha,\beta,\gamma)
= \alpha \beta +\gamma [ \alpha+\beta +\gamma -2\alpha \beta ],
\end{aligned}
\end{equation}
so that 
\begin{equation} \label{CSfull1}
(1-S(\alpha,\beta,\gamma))(1-C_{\text{out}}(\alpha,\beta,\gamma))=  (1-\alpha)(1-\beta)(1-\gamma)
\end{equation}
while $\det[H_{\text{typ}}]= -2.729 (1-\alpha)(1-\beta)(1-\gamma)$, which is indeed proportional to \eqref{CSfull1}.

\subsection{Limitations for larger QHC gates}
If we try to generalize the above approach to the full $n$-adder gate, we now have $2n+1$ inputs (2 strings of $n$ bits, $\alpha_1,\ldots,\alpha_n$ and $\beta_1,\ldots,\beta_n$, and a carry $\gamma$), and $n+1$ outputs. To estimate the complexity, let us consider the case where the block $C$ in \eqref{matrixH} is a generalization of \eqref{eq:hamiltonian8x8}, with each input bit in a $2\times 2$ matrix. Then $C$ is of size $2(2n+1)$ and $B$ is of size $n+1$. The vectors $u$ and $v$ are now replaced by $n+1$ vectors $u(1),u(2),\ldots,u(n+1)$, so there are in total $2(2n+1)(n+1)$ free parameters in the $u(i)$. We define as before the $(n+1)\times 2(2n+1)$ matrix $A$ as the matrix whose columns are the vectors $u(i)$.

Let us consider the case $n=2$. Input strings are $\alpha_1,\alpha_2,\beta_1,\beta_2$ and the carry $\gamma$. We have 3 output bits $(\mu_1\mu_2\mu_3)$, which can take 8 values from $000$ to $111$. \eqref{defS} for the $3\times 3$ matrix $S$ becomes
\begin{equation}
\label{defSnadd}
S_{\alpha_1\alpha_2\beta_1\beta_2\gamma}\left(
\begin{array}{c}\mu_1\\ \mu_2\\ \mu_3\end{array}
\right)=0.
\end{equation}
Let us first consider output $(001)$. \eqref{defSnadd} implies that the third column of $S_{\alpha_1\alpha_2\beta_1\beta_2\gamma}$ is zero when $(\alpha_1\alpha_2\beta_1\beta_2\gamma)$ is one of the inputs corresponding to output $(001)$. Thus one has for instance 
\begin{equation}
\label{Bnadd}
(B-AC_{01000}^{-1}A^T)_{i3}=0,\qquad 1\leq i\leq 3.
\end{equation}
Since $B$ is a constant matrix that does not depend on the inputs, \eqref{Bnadd} fixes the value of the $B_{i3}$ as a function of $C$ and the vectors $u(i)$, namely $B_{i3}=u(i)C_{01000}^{-1}u(3)$. The outputs $(010)$ and $(100)$ similarly fix the second and first column of $B$. Thus, these three outputs alone fix matrix $B$ entirely.

Following the analysis for $n=1$, one now needs to count the number of compatibility equations once $B$ is fixed. There are three kinds of equations:
\begin{enumerate}
\item {\em symmetry condition} of the $3\times 3$ matrix $B$: this gives 3 equations $B_{12}=B_{21}$, $B_{13}=B_{31}$, $B_{23}=B_{32}$;
\item {\em compatibility conditions for the outputs}: each output gives a constraint on part of the entries $B_{ij}$, as in  \eqref{Bnadd}. For instance output $(110)$ fixes the value $S_{i1}+S_{i2}=0$ for $1\leq i\leq 3$, which gives 3 equations. Each of the outputs $(101)$, $(011)$ and $(111)$ similarly gives 3 equations, so that there are $3\times 4=12$ such conditions;
\item {\em compatibility conditions for the inputs}: each output corresponds to several inputs which must give the same result. For instance $B_{i3}=u(i)C_{01000}^{-1}u(3)=u(i)C_{\alpha_1\alpha_2\beta_1\beta_2\gamma}^{-1}u(3)$ for any input $\alpha_1\alpha_2\beta_1\beta_2\gamma$ corresponding to output $(001)$. There are 3 such inputs, thus yielding 2 equations for each $i$. In total, 
\begin{itemize}
\item all of the 7 inputs giving $(100)$ have, for each $i$, the same value of $u(i)C_{\alpha_1\alpha_2\beta_1\beta_2\gamma}^{-1}u(1)$. This gives (for each $i$) 6 compatibility equations , thus 18 equations
\item all of the 5 inputs giving $(010)$ have, for each $i$, the same value of $u(i)C_{\alpha_1\alpha_2\beta_1\beta_2\gamma}^{-1}u(2)$ $\rightarrow 3\times 4=12$ equations
\item all of the 3 inputs giving $(001)$ have, for each $i$, the same value of $u(i)C_{\alpha_1\alpha_2\beta_1\beta_2\gamma}^{-1}u(3)$ $\rightarrow 3\times 2=6$ equations 
\item all of the 7 inputs giving $(011)$ have, for each $i$, the same value of $u(i)C_{\alpha_1\alpha_2\beta_1\beta_2\gamma}^{-1}u(2)+u(i)C_{\alpha_1\alpha_2\beta_1\beta_2\gamma}^{-1}u(3)$ $\rightarrow 3\times 6=18$ equations
\item all of the 5 inputs giving $(101)$ have, for each $i$, the same value of $u(i)C_{\alpha_1\alpha_2\beta_1\beta_2\gamma}^{-1}u(1)+u(i)C_{\alpha_1\alpha_2\beta_1\beta_2\gamma}^{-1}u(3)$ $\rightarrow 3\times 4=12$ equations
\item all of the 3 inputs giving $(110)$ have, for each $i$, the same value of $u(i)C_{\alpha_1\alpha_2\beta_1\beta_2\gamma}^{-1}u(1)+u(i)C_{\alpha_1\alpha_2\beta_1\beta_2\gamma}^{-1}u(2)$ $\rightarrow 3\times 2=6$ equations.
\end{itemize}
\end{enumerate}
In total this yields 87 equations. There are 30 free variables in the vectors $u(i)$ and 50 in $C_{00000}$ (a $10\times 10$ symmetric matrix with 10 entries used for the inputs), thus only 80 free variables. Therefore, this Hamiltonian cannot be used for higher-order gates. Note that we have only considered a very specific form for the block $C$. Therefore, the above analysis does not preclude the existence of particular $H_0$ calculating block for a specific complex logic gate.

\section{Encoding on different resonance energies at fixed quantum state}
\label{scanning}
\subsection{A new strategy}

In the preceding sections, the strategy was to measure the outputs always at the same energy (set for convenience at $E=0$). It is very convenient for applications to have all the reading blocks identical in energy but coupled to different states of the QHC calculating block. However, this is actually a quite strong condition and leads to great difficulties in extending the QHC design to more complex logic gates requiring many logical inputs. In this section we will demonstrate that instead of fixing the reading block energy to the same value for all inputs, it is possible to encode the logical outputs using the presence of a resonance in an energy interval of a reading block. As a consequence, each logical output will have a well-specified reading block identifiable by its detection energy and not by the way it is coupled to a given state of the QHC calculating block graph. This gives another freedom to the QHC logic gate optimisation problem and makes its resolution much easier. 

Let us see how this works for a half adder. We look for a calculating block described by a Hamiltonian $H_0(\alpha,\beta)$ depending on the two logical inputs $\alpha$ and $\beta$ in a symmetric way, and whose characteristic polynomial $P_{\alpha \beta}$ is such that
\begin{itemize}
\item $P_{01}$ has a root which is not a root of $P_{00}$ nor $P_{11}$;
\item $P_{11}$ has a root which is not a root of $P_{00}$ nor $P_{01}$.
\end{itemize}
For instance, if we want to measure the output of the XOR part of the half adder, we probe our system around an energy equal to the root of $P_{01}$ which is not a root of $P_{00}$ nor $P_{11}$. There will be fast oscillations when the output is 1 (input 01 or 10), but no oscillations when the output is 0 since there is no root of $P_{00}$ nor $P_{11}$ nearby.

In order to find optimal configurations fulfilling these conditions, we define the set $r_{\alpha\beta}$ of roots of $P_{\alpha \beta}$ and determine the quantities
\begin{equation}
\label{maxmin1}
\Delta_1=\max_{a_i\in r_{01}}\left(\phantom{\int}\hspace{-.4cm}\min_{b_j\in r_{00}\cup r_{11}}|a_i-b_j|\right),
\end{equation}
and 
\begin{equation}
\label{maxmin2}
\Delta_2=\max_{a_i\in r_{11}}\left(\phantom{\int}\hspace{-.4cm}\min_{b_j\in r_{00}\cup r_{01}}|a_i-b_j|\right).
\end{equation}
These two quantities determine what is the interval of maximal length around the eigenvalues which satisfy the above requirements. If these intervals are large enough, then there are energies at which only one of the outputs will be in resonance. There remains to be checked that the selected eigenvalues have a reasonable weight on the output state to be measurable. We will show examples of this procedure in the next two subsections.

\subsection{Construction of half adder Hamiltonian}

We first implement this strategy to obtain a half adder. We start by the $3\times 3$ Hamiltonian $H_0$ \eqref{eq:hamiltonianH0}. Let us consider for simplicity the case where the readout is performed on state $\ket{2}$. Our aim is to find parameters $e$, $a$ and $k$ which maximize the quantities $\Delta_1$ and $\Delta_2$ in \eqref{maxmin1}--\eqref{maxmin2}. The eigenvalues (top) with corresponding eigenvectors (bottom) of $H_0(0,0)$ are:

\begin{equation}
\label{eigensys1}
\begin{array}{ccc}
 a & e-k & e+k \\
 \{0,1,0\} & \{-1,0,1\} & \{1,0,1\} \\
\end{array}
\end{equation}

while for $H_0(1,1)$ they are:

\begin{equation}
\label{eigensys2}
\begin{array}{ccc}
 e-k & \frac{1}{2} \left(a+e+k-F\right) & \frac{1}{2} \left(a+e+k+F\right)
   \\
 \{-1,0,1\} & \left\{1,\frac{-a+e-3 k+F}{k(-k+a-e+F)-2},1\right\} &
   \left\{1,\frac{a-e+3 k+F}{k(k-a+e+F)+2},1\right\} \\
\end{array}
\end{equation}
with $F=\sqrt{a^2-2 (e+k) a+e^2+k^2+2 e k+8}$. This indicates that for $H_0(0,0)$ there is only one eigenvector having a nonzero component on state $\ket{2}$ (at $\lambda_{0,0}=a$). For  $H_0(1,1)$
there are two eigenvectors having nonzero component on state $\ket{2}$ (at $\lambda_{1,1}= \frac{1}{2} \left(a+e+k \pm F\right)$). Maximizing $|\lambda_{0,0}-\lambda_{1,1}|$ gives $a=-e$.
 By a shift in energy we can move the central energy level of $H_0$ to 0. From the same numerical maximizations for $|\lambda_{0,1}-\lambda_{1,1}|$, we find that the structural parameters $k=-2e$ are the choices that maximize $\Delta_1$ and $\Delta_2$. The corresponding Hamiltonian reads
\begin{equation}
\label{hadder2s}
H_0(\alpha,\beta)= \left[ 
    \begin{matrix}
      2e & \alpha & -2e  \\
          \alpha &0 &  \beta \\
            -2e & \beta & 2e
      \end{matrix}
    \right].      
\end{equation}

Following an optimization procedure explained in the Appendix \ref{app3}, we find that the best choice of parameters to optimize the components on the reading state is for $e=0$, with reading energies $E_1=\pm\sqrt{2}, E_2= \pm 1$ for AND and XOR respectively. We note that the characteristic polynomial of Eq~(\refeq{hadder2s}) reads 
\begin{equation}
P_{\alpha \beta} (E) =E \left(\alpha ^2+\beta ^2-E^2\right).
\label{chpolyhadder}
\end{equation}
and $P_{\alpha \beta} (E)$ at these energies is proportional to the non-Boolean determinants of AND and XOR of Table 1.

For this choice of $e=0$, the eigensystem of $H_0(\alpha,\beta, e=0)$ becomes

\begin{align}
\label{halfsimple00}
&\left\{ \begin{array}{lcl}
\lambda_1 (0,0) =0,& \,\,\,
& V_1 (0,0)= \{0, 0, 1\} \\ 
\lambda_2 (0,0) =0, & \,\,\,
& V_2 (0,0)= \{0, 1, 0\}  \\
\lambda_3 (0,0) =0,& \,\,\, 
& V_3 (0,0)= \{1, 0, 0\} 
\end{array}\right.\\
&\left\{ \begin{array}{lcl}
\lambda_1 (1,1) =-\sqrt{2},& \,\,\, 
& V_1 (1,1)=\frac{1}{2} \{1, -\sqrt{2}, 1\} \\ 
\lambda_2 (1,1) =\sqrt{2},& \,\,\, 
& V_2 (1,1)=\frac{1}{2} \{1, \sqrt{2}, 1\}  \\
\lambda_3 (1,1) =0,& \,\,\,
& V_3 (1,1)= \frac{1}{\sqrt{2}}\{-1, 0, 1\} 
\end{array}\right.\\
&\left\{ \begin{array}{lcl}
\lambda_1 (0,1) =-1,& \,\,\,
& V_1 (0,1)=\frac{1}{2} \{0, -\sqrt{2}, \sqrt{2}\}   \\ 
\lambda_2 (0,1) =1,  & \,\,\, 
& V_2 (0,1)= \frac{1}{2} \{0, \sqrt{2}, \sqrt{2}\}   \\
\lambda_3 (0,1) =0, & \,\,\,
& V_3 (0,1)= \{1, 0,0\}  
\end{array}\right.
\label{halfsimple01}
\end{align}

 \begin{figure}[!t]
    \centering
    \includegraphics[width=\linewidth]{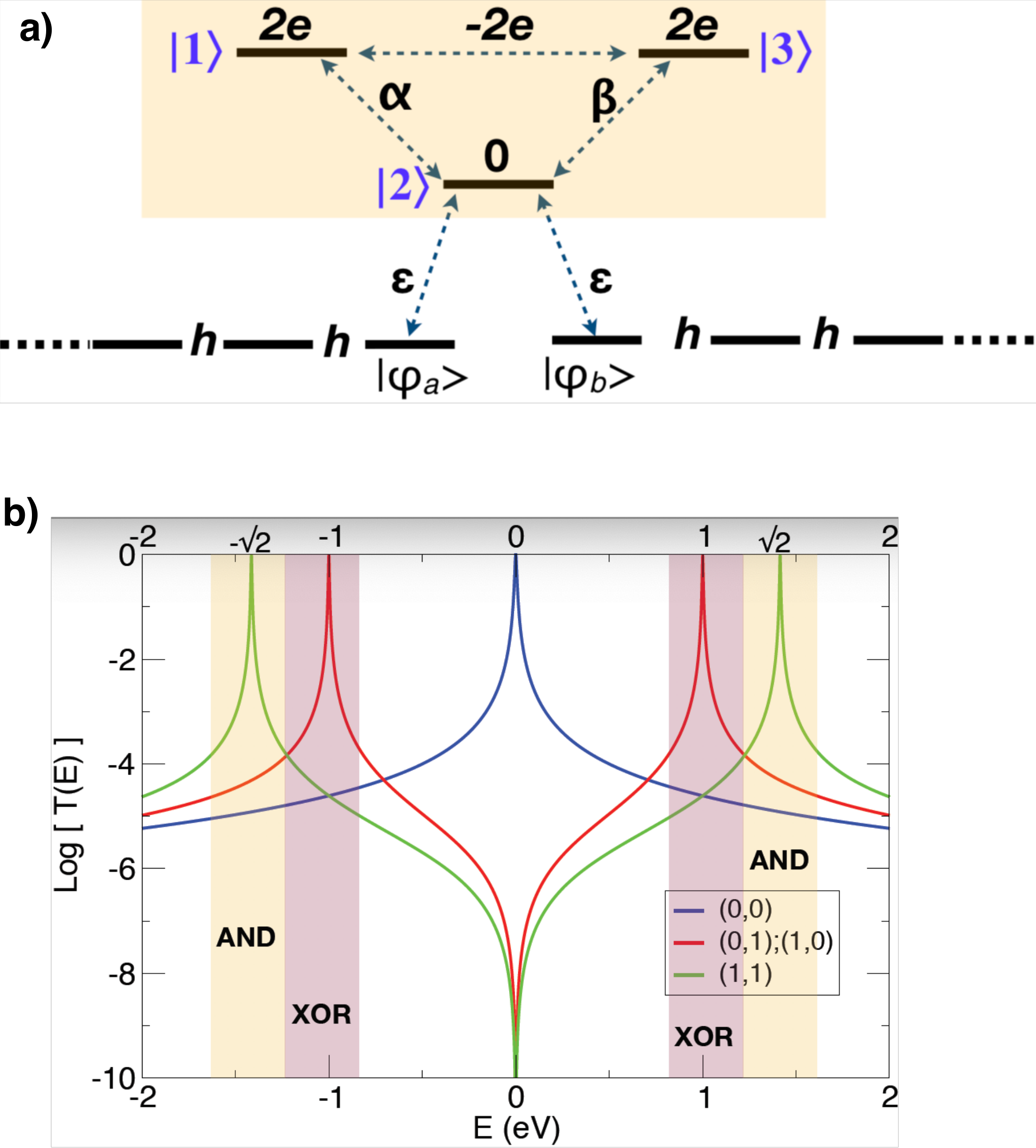} 
   \caption{a) Measurement setup of the $3 \times 3$ half adder Hamiltonian of Eq.~(\ref{hadder2s}) and b) Transmission coefficient of the Hamiltonian $H_0(\alpha,\beta)$ at $e=0$ calculated using the standard scattering theory \cite{sautet}; the measuring electrodes are here simple 1D semi-infinite tight-binding chain with inter state electronic coupling along the chains $h=4 $ eV. $\ket{\varphi_a}$ and $\ket{\varphi_b}$ are the end states of those two chains coupled to state $\ket{2}$  of the QHC gate, with coupling $\varepsilon = 0.1$ eV between $\ket{2}$ and $\ket{\varphi_a}$, $\ket{\varphi_b}$. Shaded areas correspond to the domains of energy where the reading is unambiguous.}
  \label{fig:transHA}
  \end{figure}

In order to check the efficiency of the gates for this second strategy of reading the logical outputs, we calculate the energy-dependent electronic transmission coefficient T(E) through the state $\ket{2}$ of the $H_0(\alpha,\beta)$. For this purpose, two semi-infinite tight-binding 1D chains of quantum states are connected to $\ket{2}$ via the small electronic coupling $\varepsilon$. The pointer states $\ket{\varphi_a}$ and $\ket{\varphi_b}$ are the end states of those 1D chains coupled to $\ket{2}$. T(E) is calculated using the standard scattering theory \cite{sautet} calculating the scattering matrix for a wave propagating along the 1D chains (Fig.~\ref{fig:transHA} a)) and scattered by the QHC logic gates. Fig.~\ref{fig:transHA} b) presents the T(E) variations as a function of the eigenenenergy E of such a wave determined far away from $\ket{2}$ for $e=0$. An unambiguous half adder gate is obtained with the XOR T(E) resonances at 2 energies on both sides of E=0 together with 2 others resonances for the AND. In Sec \ref{generalization} and for a fixed reading energy output strategy, the half adder needed a $5\times 5$ calculating block. Distributing the reading over one energy per output (here the XOR and the AND) leads to a minimal $3\times 3$ calculating block for a Boolean half adder, the AND and XOR output being calculated in parallel.

\subsection{Construction of full adder Hamiltonian}
We can then follow the same procedure for the 3-inputs/2-outputs Boolean full adder. We checked that a four-state calculating block is too small, as there are now 8 possible configurations of the Hamiltonian (one for each input string) with 4 energy levels each,
 and there are not enough free parameters to obtain a solution. 
We thus start with  a $5\times 5$ Hamiltonian, where we impose the reading for both outputs to be performed on level $\ket{5}$ but at different energies. We choose a $5\times 5$ Hamiltonian such that  $H(0,1,0)\equiv H(1,0,0)$ and $H(0,1,1)\equiv H(1,0,1)$. As this $5\times 5$ Hamiltonian has three more free parameters compared to the $4\times 4$  Hamiltonian,  we choose for simplicity to decouple the new fifth state from the symmetric $\ket{1}$ and $\ket{3}$ states, and also from the state $\ket{2}$. The new additional fifth state $\ket{5}$ is just then coupled to state $\ket{4}$. The $5\times 5$ Hamiltonian then reads

\begin{equation} 
\label{H5by5full}
H_{0}(\alpha,\beta,\gamma)=
\left(
\begin{array}{ccccc}
 e & \alpha  & \gamma  & k & 0 \\
 \alpha  & d & \beta  & x & 0 \\
 \gamma  & \beta  & e & k & 0 \\
 k & x & k & a & \eta  \\
 0 & 0 & 0 & \eta  & b \\
\end{array}
\right)   
\end{equation}

\begin{figure}[!t]
    \centering
    \includegraphics[width=0.6\linewidth]{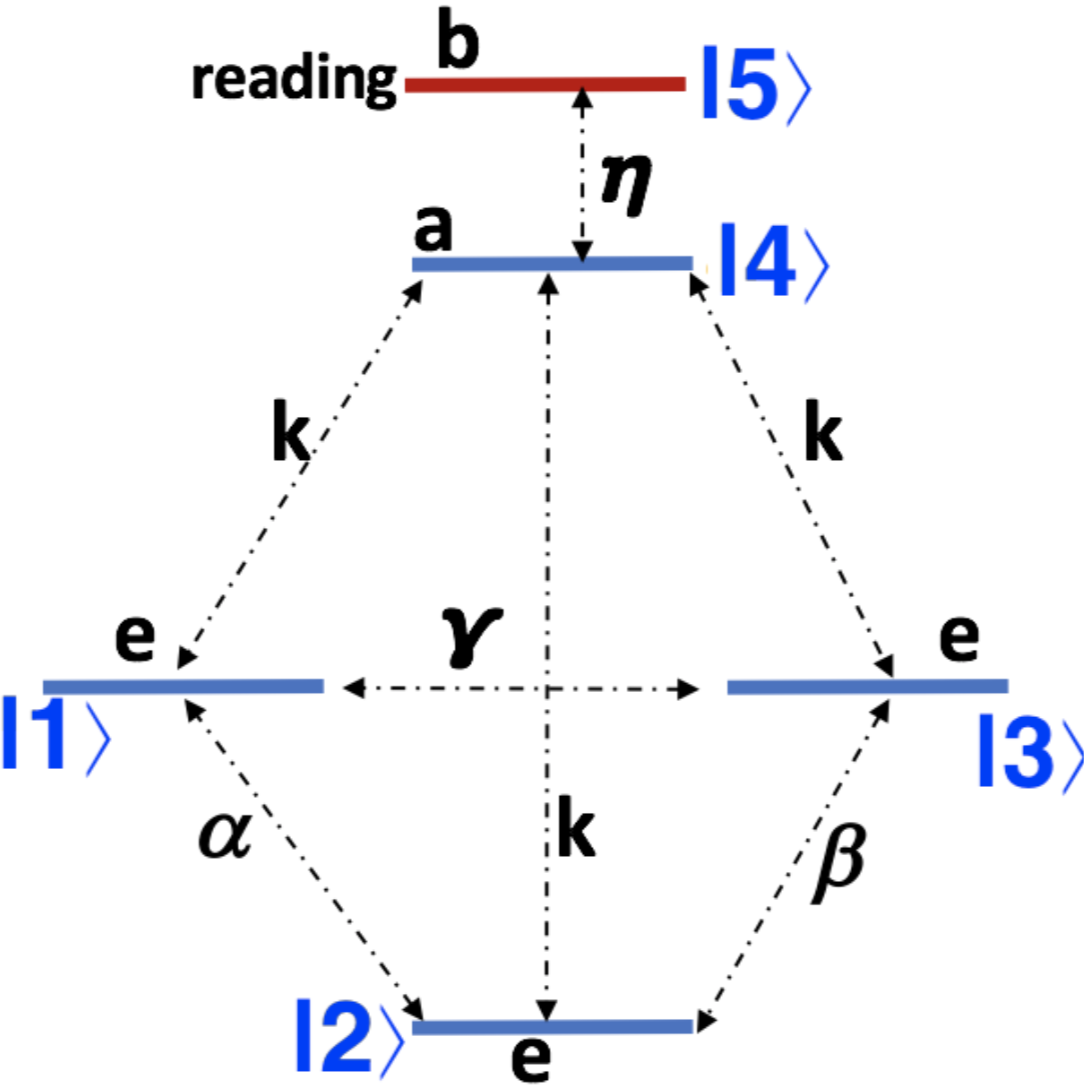} 
   \caption{A $5\times 5$ quantum graph of a 1-bit Boolean full adder. }
  \label{fig:graph_full2}
  \end{figure}

The full eigenvalues and eigenvectors calculation of the Hamiltonian Eq.~\eqref{H5by5full} shows that for $H_{0}(1,1,1)$ at eigenvalue $E=e-1$ the corresponding eigenvector has no components on states $\ket{2}, \ket{4}$and$\ket{5}$. We can shift the diagonal part of $H_{0}(\alpha,\beta,\gamma)$ by $e-1$, implying that the new onsite energies for the states $\ket{1}, \ket{3}$ will be unity ($e_1=e_3=1$). We then proceed to find conditions on the structural parameters such that eigenvalues verify  $\lambda(0,0,1)=\lambda(1,0,0)=\lambda(0,1,0)=\lambda(1,1,1)$ for output $S$ and $\lambda(0,1,1)=\lambda(1,1,0)=\lambda(1,0,1)=\lambda(1,1,1)$ for output $C_{\text{out}}$, following the truth table of Fig. (\ref{fig:fulltruth}). This leads to $d=1, x=k, b=\frac{\sqrt{3} \sqrt{\left(2 k^2+3\right) \left(-16 \eta ^2+6 k^2+9\right)}+6 k^2+9}{8 k^2+12}$. We finally maximize $|\lambda(0,1,1)-\lambda(1,0,0)|$, leading to  $a=-3/2+4  \eta^2$,$e=d=1$, $b=3/4$ and $k=(\frac{16 \eta^2 -9 }{6})^{\frac12}$. The Hamiltonian corresponds to the graph in Fig.~\ref{fig:graph_full2} with the two different reading energies being $E_1=3/2$ and $E_2=0$.
The corresponding Hamiltonian $H_0(\alpha,\beta,\gamma)$ is given by
\begin{equation} 
\label{H0full}
\left[ 
    \begin{matrix}
     1 & \alpha & \gamma  & \sqrt{  \frac{16 \eta^2 -9 }{6}}   & 0 \\
          \alpha & 1 &  \beta&  \sqrt{  \frac{16 \eta^2 -9 }{6}}  &0 \\
            \gamma & \beta & 1  &  \sqrt{  \frac{16 \eta^2 -9 }{6}}    &  0\\
            \sqrt{  \frac{16 \eta^2 -9 }{6}}   &  \sqrt{  \frac{16 \eta^2 -9 }{6}}    &  \sqrt{  \frac{16 \eta^2 -9 }{6}}   & -\frac{3}{2}+4  \eta^2 & \eta \\
            0&0&0&\eta&\frac{3}{4}
      \end{matrix}
    \right].      
\end{equation}

\begin{figure}[!t]
    \centering
    \includegraphics[width=.9\linewidth]{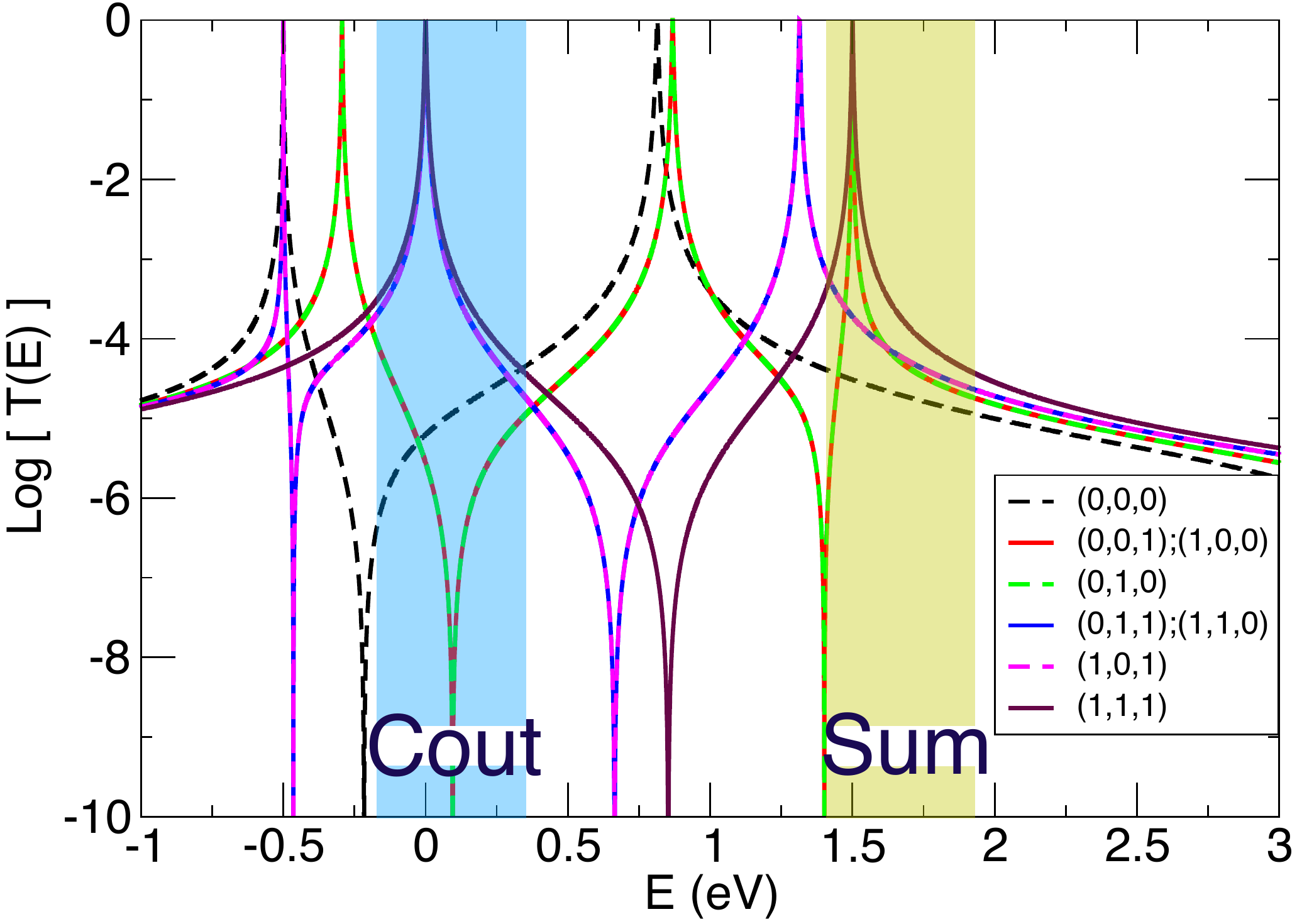} 
    \caption{Transmission coefficients of the Hamiltonian $H_0(\alpha,\beta,\gamma)$ of \eqref{H0full} with $\eta=\sqrt{15}/4$ calculated using the standard scattering theory \cite{sautet}, for different values of $(\alpha,\gamma,\beta)$. The measuring electrodes are here 1D semi-infinite tight-binding chains with inter state electronic coupling along the chains  $h=4 $ eV. $\ket{\varphi_a}$ and $\ket{\varphi_b}$ are the end states of those 2 chains coupled to state $\ket{5}$  with energy $b$ (in red in Fig.~\ref{fig:graph_full2}) of the QHC gate, with coupling $\varepsilon = 0.1$ eV between $\ket{5}$ and $\ket{\varphi_a}$, $\ket{\varphi_b}$. Shaded areas correspond to the domains of energy where the reading is unambiguous.}
  \label{fig:fulladder2st}
  \end{figure}

The characteristic polynomial of Hamiltonian Eq.~(\ref{H0full}) at $E_1=3/2$ reads:
\begin{equation} \label{Sfull2}
\frac{1}{8} \left(16 \eta ^2-9\right)(\alpha +\beta + \gamma +2(2\alpha \beta \gamma -\alpha \beta -\alpha \gamma -\gamma \beta)),
\end{equation}
which is proportional to $S(\alpha, \beta,\gamma)$ of Eq (\ref{Sfull}).
At $E_2=0$ it reads:
\begin{equation} 
\label{H0full_poly1}
\frac{1}{4} \left(16 \eta ^2-9\right) (1-\alpha ) (1-\beta ) (1-\gamma ),
\end{equation}
which is proportional to Eq (\ref{CSfull1}).
The corresponding transmission coefficient is displayed in Fig.~\ref{fig:fulladder2st}.

Interestingly enough, our second output reading strategy enables to reduce the complexity of the matrices involved to build a quantum gate: compared to the first strategy of the preceding sections (output reading at fixed energy), the half adder and full adder involve smaller number of states, and actually reach the minimal value that is needed in the QHC approach for these gates.

The protocol followed in the construction of the half adder and full adder used Eq.\eqref{maxmin1}--\eqref{maxmin2}, but made also use of the fact that for simple enough small matrices an exact analytical diagonalization is possible. For more general gates, it will be necessary to optimize numerically calculated eigenvalues and eigenvectors. The method developed in this Section is nevertheless much easier to generalize to gates with more inputs and outputs than the fixed energy strategy of the preceding section. Indeed, for an arbitrary $n$-bit gate, the eigenvalues have to fulfill the requirements derived from the truth table of the gate. Then, for a given output, one searches for intervals on which the input strings giving output 0 yield no eigenvalue, while the ones giving output 1 yield at least one eigenvalue.  

In the cases of the half adder and the adder, it was possible to find one energy value for each output satisfying the requirements.  In the general case, the protocol has no reason to single out a specific value, and will rather produce a whole range of values. It is actually easy to show that there will always be a solution, in the form of a collection of subintervals on which eigenvalues corresponding to output 1 are present and all eigenvalues for output 0 are absent. The physical procedure will therefore be to scan these subintervals: output 1 will correspond to presence of a resonance in one of the subintervals.

In general this produces solutions of the problem for arbitrary complicated gates. Nevertheless the number of subintervals will grow with gate complexity, and their size will decrease. To remain within experimental precision, careful search for the largest subintervals should be made. The result is not easy to predict: while for generic values of the parameters, the average size of subintervals is bound to decrease on with the density of states, the optimal value is much more difficult to evaluate, and requires a precise mathematical study of extreme value distributions to predict its asymptotic behaviour for complex $n$-bit gates.

\section{Conclusion}

In this paper, we have presented different strategies and protocols to build complex gates within the QHC approach. We have reinterpreted previous results and shown that one can build simple gates such as the half adder by merging in a specific way the Hamiltonian matrices of elementary gates such as XOR and AND. This paves the way to a systematic way of building computational Hamiltonians from simple gates. We then presented an explicit protocol to construct Hamiltonian matrices performing arbitrary complex operations in the fixed energy approach. This enabled us to produce the first explicit example of full adder in this framework. However, using the protocol we built, we also show that to construct more complex gates will become extremely difficult. We therefore proposed a second design, where different energies are used for the readout. We presented a protocol to construct Hamiltonian matrices with this new design, and used it to construct an explicit half adder and full adder. We also discussed the possibility of more complex gates, showing that within this approach there should be better possibilities to access the limit of large number of inputs and outputs.

Our results show that with Quantum Hamiltonian Computing approach, it is possible to produce explicit Hamiltonians for full operations such as the digital adder. Within the second presented output reading strategy, more complex operations can be also constructed. In principle, Quantum Hamiltonian Computing is well designed for applications to single molecule electronics. The protocols presented here enable to construct whole families of solutions for a given gate. We thus think that the use of these protocols should allow to produce tailored Hamiltonians suitable for experimental molecular implementations of these gates, starting with single molecule half adders and then molecule full adders.

\section*{acknowledgments}This work was supported in part by the Programme Investissements d'Avenir ANR-11-IDEX-0002-02, reference
ANR-10-LABX-0037-NEXT (project COMPMOL) and by CNRS. Computational  resources  were  provided
by CALcul en MIdi-Pyr\'en\'ees (CALMIP).
\appendix
\appendixpage
\section{Half adder at fixed reading energy}
\label{app1}
Here we solve explicitly the equations \eqref{condB1h} obtained for the half adder,
\begin{equation}
\label{condB1hrappel}
uC_{11}^{-1}v=uC_{01}^{-1}v=uC_{10}^{-1}v,\quad vC_{01}^{-1}v=vC_{10}^{-1}v,
\end{equation}
and find a 5-parameter family of solutions for the half adder. Conditions \eqref{condB1hrappel} can be rewritten
\begin{equation}
\label{condB1h2}
u(C_{01}^{-1}-C_{11}^{-1})v=0, u(C_{10}^{-1}-C_{11}^{-1})v=0, v(C_{01}^{-1}-C_{10}^{-1})v=0.
\end{equation}
Introducing $C_{00}$ we get that there has to be real numbers $r$ and $s$ such that
\begin{align}
\label{condB1h3s}
u(C_{01}^{-1}-C_{00}^{-1})v=u(C_{10}^{-1}-C_{00}^{-1})v=u(C_{11}^{-1}-C_{00}^{-1})v&=-s,\\
 v(C_{01}^{-1}-C_{00}^{-1})v=v(C_{10}^{-1}-C_{00}^{-1})v&=-r.
 \label{condB1h3r}
\end{align}
In order to have unicity of the solution (one-dimensional kernels), the kernel of $S$ should also be one-dimensional. Since $S$ is $2\times 2$, this means that $S$ should not be the zero matrix. This leads to further conditions: the second column of $S_{11}$ should not be zero, and the first column of $S_{01}$ and $S_{10}$ should not be zero. One obtains the additional conditions 
\begin{eqnarray}
u(C_{11}^{-1}-C_{01}^{-1})u\neq 0\quad\textrm{ or}&u(C_{01}^{-1}-C_{11}^{-1})v\neq 0\\
u(C_{01}^{-1}-C_{11}^{-1})v\neq 0\quad\textrm{ or}&v(C_{01}^{-1}-C_{11}^{-1})v\neq 0,
\end{eqnarray}
which taking into account \eqref{condB1h2} yields
\begin{eqnarray}
u(C_{11}^{-1}-C_{01}^{-1})u&\neq 0\\
v(C_{01}^{-1}-C_{11}^{-1})v&\neq 0.
\end{eqnarray}
There is an additional condition associated with input $(0,0)$. From \eqref{matrixBh} we get
\begin{equation}
\label{matrixSh}
S_{00}=\left(
\begin{array}{c c}
u(C_{11}^{-1}-C_{00}^{-1})u\quad& u(C_{01}^{-1}-C_{00}^{-1})v\\
v(C_{11}^{-1}-C_{00}^{-1})u\quad& v(C_{01}^{-1}-C_{00}^{-1})v
\end{array}
\right).
\end{equation}
As before, suppose we are in the case where $C_{00}$ is invertible. Then we need $\det S_{00}\neq 0$.

We now consider in detail the case where $C$ is of the form
\begin{equation}
C_{\alpha\beta}=\left(
\begin{array}{cccc}
  &\alpha & & \\
\alpha  & & & \\
  &  & &\beta \\
  & & \beta& 
\end{array}
\right).
\end{equation}
In such a case we can apply the Woodbury identity \cite{Gollub}, which states that if M is $n\times n$, $U$ is $n\times k$, $D$ is $k\times k$ and $V$ is $k\times n$, then
\begin{equation}
\label{woodbury}
(M+UDV)^{-1}=M^{-1}-M^{-1}U(D^{-1}+VM^{-1}U)^{-1}VM^{-1}.
\end{equation}
Matrix $C$ is of the form $C_a=C_{00}+U^T_aD_aU_a$, with $D_a$ of size the rank of $C_a$. We get
\begin{equation}
\label{wood2h}
C_a^{-1}-C_{00}^{-1}=-C_{00}^{-1}U^T_a(D_a^{-1}+U_aC_{00}^{-1}U^T_a)^{-1}U_aC_{00}^{-1}.
\end{equation}
If we set $Q'=C_{00}^{-1}+C_{a}-C_{00}$, it is easy to check that $D_{a}^{-1}+U_{a}C_{00}^{-1}U_{a}^T=U_{a}Q'U_{a}^T$, so that \eqref{wood2h}  gives
\begin{equation}
\label{wood2h1}
C_a^{-1}-C_{00}^{-1}=-C_{00}^{-1}U^T_a(U_{a}Q'U_{a}^T)^{-1}U_aC_{00}^{-1}.
\end{equation}
Making the change of variables
\begin{equation}
\tilde{u}=C_{00}^{-1}u,\quad\tilde{v}=C_{00}^{-1}v,
\end{equation}
we have for instance
\begin{equation}
\label{uuh}
u(C_a^{-1}-C_{00}^{-1})v=-\tilde{u} U^T_a(U_{a}Q'U_{a}^T)^{-1}U_a\tilde{v}.
\end{equation}
One can check that 
\begin{equation}
\label{projecth1}
u(C_{10}^{-1}-C_{00}^{-1})v=- \left(\begin{array}{c}\tilde{u}_{1}\\ \tilde{u}_{2}
\end{array}
\right)
 T_{12}^{-1}\left(\begin{array}{c}\tilde{v}_{1}\\ \tilde{v}_{2}
\end{array}
\right),
\end{equation}
with
\begin{equation}
T_{12}=\left(\begin{array}{cc}
Q_{11}&Q_{12}+1\\ Q_{21}+1&Q_{22}
\end{array}
\right),\qquad Q=C_{00}^{-1},
\end{equation}
and
\begin{equation}
\label{projecth2}
u(C_{11}^{-1}-C_{00}^{-1})v=- \left(\begin{array}{c}\tilde{u}_{1}\\ \tilde{u}_{2}\\\tilde{u}_{3}\\ \tilde{u}_{4}
\end{array}
\right)T_{1234}^{-1}\left(\begin{array}{c}\tilde{v}_{1}\\ \tilde{v}_{2}\\\tilde{v}_{3}\\ \tilde{v}_{4}
\end{array}
\right)
\end{equation}
with
\begin{equation}
T_{1234}=\left(
\begin{array}{cccc}
 Q_{11} & Q_{12}+1 & Q_{13} & Q_{14} \\
 Q_{21}+1 & Q_{22} & Q_{23} & Q_{24} \\
 Q_{31} & Q_{32} & Q_{33} & Q_{34}+1 \\
 Q_{41} & Q_{42} & Q_{43}+1 & Q_{44} \\
\end{array}
\right).
\end{equation}
Therefore, conditions \eqref{condB1h3s}  and  \eqref{condB1h3r} are equivalent to
\begin{eqnarray}
\label{eqvvh}
\left(\begin{array}{c}\tilde{v}_{1}\\ \tilde{v}_{2}
\end{array}
\right)
T_{12}\left(\begin{array}{c}\tilde{v}_{1}\\ \tilde{v}_{2}
\end{array}
\right)=r,\quad &  \left(\begin{array}{c}\tilde{v}_{3}\\ \tilde{v}_{4}
\end{array}
\right)
T_{34}\left(\begin{array}{c}\tilde{v}_{3}\\ \tilde{v}_{4}
\end{array}
\right)=r,\\
\left(\begin{array}{c}\tilde{u}_{1}\\ \tilde{u}_{2}
\end{array}
\right)
 T_{12}\left(\begin{array}{c}\tilde{v}_{1}\\ \tilde{v}_{2}
\end{array}
\right)=s,\quad &  \left(\begin{array}{c}\tilde{u}_{3}\\ \tilde{u}_{4}
\end{array}
\right)
 T_{34}\left(\begin{array}{c}\tilde{v}_{3}\\ \tilde{v}_{4}
\end{array}
\right)=s,
\label{equv1h}
\end{eqnarray}
and
\begin{equation}
\label{equv2h}
\left(\begin{array}{c}\tilde{u}_{1}\\ \tilde{u}_{2}\\\tilde{u}_{3}\\ \tilde{u}_{4}
\end{array}
\right)
T_{1234}\left(\begin{array}{c}\tilde{v}_{1}\\ \tilde{v}_{2}\\\tilde{v}_{3}\\ \tilde{v}_{4}
\end{array}
\right)=s,
\end{equation}
which gives 10 variables and 5 equations. We can change $\tilde{v}$ to $\hat{v}=\tilde{v}/\sqrt{r}$ and $\tilde{u}$ to $\hat{u}=\tilde{u}\sqrt{r}/s$, so that the above equations become
\begin{eqnarray}
\label{eqvvhbis}
\left(\begin{array}{c}\hat{v}_{1}\\ \hat{v}_{2}
\end{array}
\right)
T_{12}\left(\begin{array}{c}\hat{v}_{1}\\ \hat{v}_{2}
\end{array}
\right)=1,\quad &  \left(\begin{array}{c}\hat{v}_{3}\\ \hat{v}_{4}
\end{array}
\right)
T_{34}\left(\begin{array}{c}\hat{v}_{3}\\ \hat{v}_{4}
\end{array}
\right)=1,&\\
\left(\begin{array}{c}\hat{u}_{1}\\ \hat{u}_{2}
\end{array}
\right)
 T_{12}\left(\begin{array}{c}\hat{v}_{1}\\ \hat{v}_{2}
\end{array}
\right)=1,\quad &  \left(\begin{array}{c}\hat{u}_{3}\\ \hat{u}_{4}
\end{array}
\right)
 T_{34}\left(\begin{array}{c}\hat{v}_{3}\\ \hat{v}_{4}
\end{array}
\right)=1,\label{equvhbis}\\
\left(\begin{array}{c}\hat{u}_{1}\\ \hat{u}_{2}\\\hat{u}_{3}\\ \hat{u}_{4}
\end{array}
\right)
T_{1234}\left(\begin{array}{c}\hat{v}_{1}\\ \hat{v}_{2}\\\hat{v}_{3}\\ \hat{v}_{4}
\end{array}
\right)=1.&
\label{equvhter}
\end{eqnarray}
We can express $\hat{v}_2$ and $\hat{v}_4$ as a function of $\hat{v}_1$ and $\hat{v}_3$ using  ~\eqref{eqvvhbis}, and use  \eqref{equvhbis} to eliminate $\hat{u}_2$ and $\hat{u}_4$. The remaining equation \eqref{equvhter} can be used to eliminate $\hat{u}_3$. We are left with 3 free parameters  $\hat{v}_1$,  $\hat{v}_3$ and  $\hat{u}_1$, each of which giving a solution. Moreover, each vector can be multiplied by an arbitrary factor (this corresponds to the freedom in the choice of $r$ and $s$). Thus we end up with 5 free parameters, consistent with the fact that we had started with 8 variables $u_i$ and $v_i$ and 3 compatibility equations \eqref{condB1hrappel}.

\section{Full adder at fixed reading energy}
\label{app2}
\subsection{The equations}
Here we show how to solve \eqref{rel1} to \eqref{condt6}. The steps are similar to those for the half adder. When \eqref{condt6} holds, \eqref{rel1} and \eqref{condB2} are equivalent to the existence of a real number $t$ such that
\begin{eqnarray}
\label{condt1}
&u(C_{011}^{-1}-C_{111}^{-1})u=t\\
\label{condt2}
&v(C_{001}^{-1}-C_{111}^{-1})v=t,\\
&u(C_{001}^{-1}-C_{111}^{-1})v=-t.
\label{condt3}
\end{eqnarray}
If $C$ is fixed, we want to find vectors $u$ and $v$ satisfying \eqref{condt4}--\eqref{condt6} and \eqref{condt1}--\eqref{condt3}. Moreover, in order to have unicity of the solution (one-dimensional kernels), the kernel of $S$ should also be one-dimensional. Since $S$ is $2\times 2$, this means that $S$ should not be the zero matrix. This leads to further conditions: the second column of $S_{011}$ should not be zero (i.e. $u(C_{001}^{-1}-C_{011}^{-1})v\neq 0$ or $v(C_{001}^{-1}-C_{011}^{-1})v\neq 0$), the first column of $S_{001}$ should not be zero (i.e. $u(C_{011}^{-1}-C_{001}^{-1})u\neq 0$ or $u(C_{011}^{-1}-C_{001}^{-1})v\neq 0$), and $S_{111}\neq 0$. One obtains the three additional conditions 
\begin{eqnarray}
u(C_{001}^{-1}-C_{011}^{-1})u&\neq 0\\
v(C_{001}^{-1}-C_{011}^{-1})v&\neq 0\\
t&\neq 0.
\end{eqnarray}
The final condition is that for input $(0,0,0)$. We have the identity $\det H=\det C \det S$, therefore we get
\begin{equation}
\label{matrixS}
S_{000}=\left(
\begin{array}{c c}
u(C_{011}^{-1}-C_{000}^{-1})u\quad& u(C_{001}^{-1}-C_{000}^{-1})v\\
v(C_{011}^{-1}-C_{000}^{-1})u\quad& v(C_{001}^{-1}-C_{000}^{-1})v
\end{array}
\right).
\end{equation}
There are two possibilities. Either the Hamiltonian $H$ has no eigenvector with eigenvalue 0, in which case det$C_{000}\neq 0$ and $\det S_{000}\neq 0$. Or $H_{000}$ has an eigenvector with eigenvalue 0 of the form $(p_1,\ldots,p_{N-2},0,0)$, in which case, from  \eqref{cp}, $C_{000}$ must be non-invertible (otherwise the only solution is $p=0$), so that one must have det$C_{000}=0$, and from \eqref{up}--\eqref{vp}  $u$ and $v$ must be orthogonal to the vector in the kernel of $C_{000}$.\\

\subsection{Elimination of variables}\label{sec:elimin}
Suppose we are in the case where $C_{000}$ is invertible. For any input $a=\alpha\beta\gamma$, matrices $C_a$ are of the form $C_a=C_{000}+U^T_aD_aU_a$, where matrices $U_a$ and $D_a$ only depend on the position of the input in the Hamiltonian. For instance for $C_{100}$ one has 
\begin{equation}
D_{100}=\left(\begin{array}{cc}
\frac12&0\\
0&-\frac12
\end{array}
\right)
\end{equation}
and
\begin{equation}
U_{100}=\left(\begin{array}{ccccc}
\hdots&1&\hdots&1&\hdots\\
\hdots&1&\hdots&-1&\hdots
\end{array}
\right)
\end{equation}
where dots stand for zero entries, and nonzero entries are located at positions corresponding to the position of $\alpha$. All equations \eqref{condt1}--\eqref{matrixS} can be written in terms of differences $C_a^{-1}-C_{000}^{-1}$. 
Using the Woodbury formula \eqref{woodbury} we get
\begin{equation}
\label{wood2}
C_a^{-1}-C_{000}^{-1}=-C_{000}^{-1}U^T_a(D_a^{-1}+U_aC_{000}^{-1}U^T_a)^{-1}U_aC_{000}^{-1}.
\end{equation}
Making the change of variables $\tu=C_{000}^{-1}u,\quad\tv=C_{000}^{-1}v$, we have
\begin{equation}
\label{uv}
u(C_a^{-1}-C_{000}^{-1})v=-\tu\left[ \phantom{\int}\!\!\!\!\!U^T_a(D_a^{-1}+U_aC_{000}^{-1}U^T_a)^{-1}U_a\right]\tv.
\end{equation}
This quantity depends on $\tu$ and $\tv$ only via $U_a\tu$ and $U_a\tv$. Matrices $U_a$ have nonzero components only at indices where inputs are. For instance if $H_{12}=H_{21}=\alpha$, then $U_{100}\tu$ only depends on $\tu_1$ and $\tu_2$. Reexpressing all equations in terms of the new variables $\tu$ and $\tv$ (since $u$ and $v$ are free this is still completely general), we will get much simpler systems.

\subsection{Particular case}
We consider from now on the case where $C$ is of the form
\begin{equation}
C_{\alpha\beta}=\left(
\begin{array}{cccccc}
  &\alpha & & & &\\
\alpha  & & & & &\\
  &  & &\beta & &\\
  & & \beta& & &\\
  & & & & & \gamma\\
  & & & & \gamma & 
\end{array}
\right).
\end{equation}
One then has the form $C_a=C_{000}+U^T_aD_aU_a$ with for instance for $C_{100}$
\begin{equation}
U_{100}=\left( \begin{array}{cccc}
1&1&0&\hdots\\
1&-1&0&\hdots
\end{array}
\right),\quad D_{100}=\left(\begin{array}{cc}
\frac12&0\\
0&-\frac12
\end{array}
\right).
\end{equation}
Defining $Q=C_{000}^{-1}$ and $Q'=C_{000}^{-1}+C_{111}-C_{000}$, one can easily check that $D_{a}^{-1}+U_{a}QU_{a}^T=U_{a}Q'U_{a}^T$. Thus \eqref{uv} becomes
\begin{equation}
\label{uCav}
u(C_{a}^{-1}-C_{000}^{-1})v=-\tu U_{a}^T (U_{a}Q'U_{a}^T)^{-1}U_{a}\tv.
\end{equation}
Matrix $U_{a}$ projects $Q'$ and $\tv$ over the subspace where it is nonzero. One can check by direct calculation that the following identities hold:
\begin{equation}
\label{equv100}
\tu U_{100}^T (U_{100}Q'U_{100}^T)^{-1}U_{100}\tv=
\left(\begin{array}{c}\tu_1\\ \tu_2
\end{array}
\right)T_{12}\left(\begin{array}{c}\tv_1\\ \tv_2
\end{array}
\right)
\end{equation}
\begin{equation}
\label{equv110}
\tu U_{110}^T (U_{110}Q'U_{110}^T)^{-1}U_{110}\tv=\left(\begin{array}{c}\tu_1\\ \tu_2\\ \tu_3\\ \tu_4
\end{array}
\right)T_{1234}\left(\begin{array}{c}\tv_1\\ \tv_2\\ \tv_3\\ \tv_4
\end{array}
\right)
\end{equation}
\begin{equation}
\tu U_{111}^T (U_{111}Q'U_{111}^T)^{-1}U_{111}\tv=\tu R \tv,
\label{equv111}
\end{equation}
with
\begin{equation}
T_{12}=\left(
\begin{array}{cccc}
 Q_{11} & Q_{12}+1 \\
 Q_{21}+1 & Q_{22} \end{array}
\right)^{-1}
\end{equation}
and
\begin{equation}
T_{1234}=\left(
\begin{array}{cccc}
 Q_{11} & Q_{12}+1 & Q_{13} & Q_{14} \\
 Q_{21}+1 & Q_{22} & Q_{23} & Q_{24} \\
 Q_{31} & Q_{32} & Q_{33} & Q_{34}+1 \\
 Q_{41} & Q_{42} & Q_{43}+1 & Q_{44} \\
\end{array}
\right)^{-1},
\end{equation}
and similarly for other indices, and $R=Q'^{-1}$.\\

\subsection{The equations}
There are 11 equations \eqref{rel1}--\eqref{condt6}. They can be rewritten in terms of $\tu$ and $\tv$. Equations \eqref{condt4} read $u(C_{011}^{-1}-C_{000}^{-1})u= u(C_{101}^{-1}-C_{000}^{-1})u=u(C_{110}^{-1}-C_{000}^{-1})u=-q$, which gives
\begin{equation}
\label{uQu}
\left(\begin{array}{c}\tu_1\\ \tu_2\\ \tu_3\\ \tu_4
\end{array}
\right)T_{1234}\left(\begin{array}{c}\tu_1\\ \tu_2\\ \tu_3\\ \tu_4
\end{array}
\right)=q
\end{equation}
and two similar equations for $T_{1256}$ and $T_{3456}$. Equations \eqref{condt5} read $v(C_{001}^{-1}-C_{000}^{-1})v=v(C_{010}^{-1}-C_{000}^{-1})v=v(C_{100}^{-1}-C_{000}^{-1})v=-r$ , which gives
\begin{equation}
\label{vQv}
\left(\begin{array}{c}\tv_1\\ \tv_2
\end{array}
\right)T_{12}\left(\begin{array}{c}\tv_1\\ \tv_2
\end{array}
\right)=r
\end{equation}
and two similar equations for $T_{34}$ and $T_{56}$. 
Equations \eqref{condt6} read $u(C_{001}^{-1}-C_{000}^{-1})v= u(C_{010}^{-1}-C_{000}^{-1})v=u(C_{100}^{-1}-C_{000}^{-1})v=-s$ and $u(C_{011}^{-1}-C_{000}^{-1})v= u(C_{101}^{-1}-C_{000}^{-1})v=u(C_{110}^{-1}-C_{000}^{-1})v=-s$, that is,
\begin{equation}
\label{equv}
\left(\begin{array}{c}\tu_1\\ \tu_2
\end{array}
\right)T_{12}\left(\begin{array}{c}\tv_1\\ \tv_2
\end{array}
\right)=s,\quad
\left(\begin{array}{c}\tu_1\\ \tu_2\\ \tu_3\\ \tu_4
\end{array}
\right)T_{1234}\left(\begin{array}{c}\tv_1\\ \tv_2\\ \tv_3\\ \tv_4
\end{array}
\right)=s
\end{equation}
and four similar equations. Since from \eqref{uCav} and \eqref{equv111} one has $u(C_{111}^{-1}-C_{000}^{-1})v=-\tu R\tv$, \eqref{condt1}--\eqref{condt3} can be rewritten
\begin{align}
\tu R \tu&=q+t\label{uRu}\\
\tv R \tv&=r+t\label{vRv}\\
\tu R \tv&=s-t\label{uRv}.
\end{align}
We have moved from 11 to 15 equations \eqref{uQu}--\eqref{uRv}, at the expense of 4 additional parameters $q,r,s,t$.
Finally, matrix $S_{000}$ in \eqref{matrixS} can be rewritten as
\begin{equation}
\label{matrixS2}
S_{000}=-\left(
\begin{array}{c c}
q& s\\
s& r\end{array}
\right),
\end{equation}
 so that the constraint $\det S_{000}\neq 0$ is simply equivalent to $qr-s^2\neq 0$.

 \subsection{Solving the equations}
We first change variables $\tu$ to $\hu=\tu \sqrt{r}/s$ and $\tv$ to $\hv=\tv/\sqrt{r}$, so that  \eqref{vQv} becomes
\begin{equation}
\label{equvbis2}
\left(\begin{array}{c}\hv_1\\ \hv_2\
\end{array}
\right)T_{12}\left(\begin{array}{c}\hv_1\\ \hv_2
\end{array}
\right)=1, 
\end{equation}
and \eqref{equv} yield
\begin{equation}
\label{equv2}
\left(\begin{array}{c}\hu_1\\ \hu_2
\end{array}
\right)T_{12}\left(\begin{array}{c}\hv_1\\ \hv_2
\end{array}
\right)=1,\quad
\left(\begin{array}{c}\hu_1\\ \hu_2\\ \hu_3\\ \hu_4
\end{array}
\right)T_{1234}\left(\begin{array}{c}\hv_1\\ \hv_2\\ \hv_3\\ \hv_4
\end{array}
\right)=1.
\end{equation}
Equations \eqref{uQu} gives
\begin{equation}
\label{uQu2}
\left(\begin{array}{c}\hu_1\\ \hu_2\\ \hu_3\\ \hu_4
\end{array}
\right)T_{1234}\left(\begin{array}{c}\hu_1\\ \hu_2\\ \hu_3\\ \hu_4
\end{array}
\right)=\lambda,
\end{equation}
with $\lambda=q r/s^2$. Equations \eqref{uRu}--\eqref{uRv} become
\begin{align}
\hu R \hu&=\lambda+\frac{(1-\sigma)^2}{\tau-1}\label{uRu2}\\
\hv R \hv&=\tau\label{vRv2}\\
\hu R \hv&=\sigma,
\label{uRv2}
\end{align}
with $\sigma=1-t/s$, and $\tau=1+t/r$. Condition on $S_{000}$ below \eqref{matrixS2} amounts to $\lambda^2\neq 1$.\\

We can first solve \eqref{equvbis2}, expressing $\hv_{2k}$ as a function of $\hv_{2k-1}$, for $1\leq k \leq 3$. The 6 equations \eqref{equv2} in $\hu$ then allow to express $\hu_1,\ldots,\hu_6$ as a function of $\hv$. Inverting the relations between constants yields 
\begin{equation}
\label{qrs}
q=\lambda\frac{\tau-1}{(1-\sigma)^2}t, \qquad r=\frac{1}{\tau-1}t,\qquad s=\frac{1}{1-\sigma}t.
\end{equation}
The fact that all constants are proportional to $t$ means that once a solution has been found, vectors $u$ and $v$ can always be multiplied by some constant (which corresponds to the fact that our original equations were homogeneous quadratic equations). Hence one of the parameters just corresponds to this scaling freedom, and we are left with 6 equations \eqref{uQu2}--\eqref{uRv2} and 6 parameters, $v_1, v_3, v_5$ and $\lambda,\sigma, \tau$. The values of $\lambda$, $\sigma$ and $\tau$ in terms of $v_1, v_3, v_5$ are readily obtained from \eqref{uRu2}--\eqref{uRv2}. Injecting these expressions into \eqref{uQu2} yields 3 remaining equations
\begin{equation}
\label{eqfin}
\left(\begin{array}{c}\hu_1\\ \hu_2\\ \hu_3\\ \hu_4
\end{array}
\right)T_{1234}\left(\begin{array}{c}\hu_1\\ \hu_2\\ \hu_3\\ \hu_4
\end{array}
\right)=\hu R \hu+\frac{(1-\hu R \hv)^2}{1-\hv R \hv}
\end{equation}
depending on 3 remaining variables $v_1, v_3, v_5$. These equations can then be solved numerically for given entries of the matrix $C$.

\section{Half adder with different reading energies}
\label{app3}

We start from the Hamiltonian \eqref{hadder2s} which already maximizes the quantities $\Delta_1$ and $\Delta_2$ in \eqref{maxmin1}--\eqref{maxmin2}. The eigenvalues $\lambda_i (\alpha,\beta)$ and eigenvectors $V_i (\alpha,\beta)$ of $H_0(\alpha,\beta)$, for $1\leq i\leq 3$, are given by
\begin{align} \label{half00}
&\left\{ 
\begin{array}{ll}
\lambda_1 (0,0) =4e \,,
& V_1 (0,0)=\frac{1}{\sqrt{2}} \{-1, 0, 1\} \\ 
\lambda_2 (0,0) =0 \,, 
& V_2 (0,0)=\frac{1}{\sqrt{2}} \{1, 0, 1\}  \\
\lambda_3 (0,0) =0 \,, 
& V_3 (0,0)= \{0, 1, 0\} 
\end{array}\right.\\
&\left\{ \begin{array}{ll}
\lambda_1 (1,1) =-\sqrt{2}\,, 
& V_1 (1,1)=\frac{1}{2} \{1, -\sqrt{2}, 1\} \\ 
\lambda_2 (1,1) =\sqrt{2}\,, 
& V_2 (1,1)=\frac{1}{2} \{1, \sqrt{2}, 1\}  \\
\lambda_3 (1,1) =4e\,, 
& V_3 (1,1)= \frac{1}{\sqrt{2}}\{-1, 0, 1\} 
\end{array}\right.\\
&\left\{ \begin{array}{ll}
&\lambda_1 (0,1) =\frac{16 e^2+4 e f+f^2+3}{3 f} \,, \\
& V_1 (0,1)= \left\{\frac{2 e \lambda_1-\lambda_1^2+1}{2 e \lambda_1},\frac{1}{\lambda_1},1\right\}  \\ 
&\lambda_2 (0,1) = -\frac{16 e^2 z-8 ef+f^2 \bar{z}+3 z}{6 f}  \,, \\
& V_2 (0,1)= \left\{\frac{2 e \lambda_2-\lambda_2^2+1}{2 e \lambda_2},\frac{1}{\lambda_2},1\right\}  \\
&\lambda_3 (0,1) =\frac{16 e^2 \bar{z}-8 e f+f^2 z+3 y}{6 f} \,, \\
& V_3 (0,1)=  \left\{\frac{2 e \lambda_3-\lambda_3^2+1}{2 e \lambda_3},\frac{1}{\lambda_3},1\right\} 
\end{array}\right.\,
\label{half01}
\end{align}
where $f=(64 e^3+3 \sqrt{3} \sqrt{-128 e^4-13e^2-1}-9e)^{\frac13}$, $z=1+i \sqrt{3}$ and $\bar{z}=1-i \sqrt{3}$. From \eqref{half00}--\eqref{half01} it is clear that by measuring at the energies $E_1=\pm\sqrt{2}$ and probing state $\ket{2}$ we can obtain the AND gate (note that for third eigenvalue of AND gate with $\lambda_3(1,1)=4e$ there is no component on state $\ket {2}$), with a component of the eigenvector equal to $\pm\sqrt{2}/2$. For the $H_0(0,0)$ case there is only one eigenvector at $E=0$ with a nonzero component on state $\ket{2}$. In order to have symmetric eigenvalues with maximum splitting and equal components on state $\ket{2}$ one can optimise Eq.~(\ref{half01}) to find the value of the parameter $e$ such that the weight on $\ket{2}$ is also $\pm\sqrt{2}/2$ when the XOR gate is measured; this leads to $e=0$.


\bibliographystyle{aip}

\end{document}